\documentclass[preprint,singlecolumn,superscriptaddress,preprintnumbers,nopacs,floatfix,amsmath,amssymb]{revtex4}
\newcommand{\lbl}[1]{\label{#1}}

\usepackage{graphicx}
\usepackage{dcolumn}
\usepackage{bm}
\usepackage{amsmath}
\usepackage{amssymb}
\usepackage{mathrsfs}
\usepackage{bbm}
\usepackage{epsfig}

\usepackage{wrapfig}
\usepackage{eurosym}
\usepackage[usenames,dvipsnames]{color}
\usepackage[yyyymmdd,hhmmss]{datetime}

\newcommand{\be}{\begin{eqnarray}}
\newcommand{\ee}{\end{eqnarray}}

\newcommand{\eastar}{\end{eqnarray*}}

\begin{document}

\title{ GAUGE FIELDS, STRINGS, SOLITONS, ANOMALIES, \\  ~ \\ AND  THE SPEED OF LIFE  \\ ~ \\ }

\author{Antti J. Niemi}
\email{Antti.Niemi@physics.uu.se}
\affiliation{
Laboratoire de Mathematiques et Physique Theorique
CNRS UMR 6083, F\'ed\'eration Denis Poisson, Universit\'e de Tours,
Parc de Grandmont, F37200, Tours, France}\affiliation{Department of Physics and Astronomy, Uppsala University,
P.O. Box 803, S-75108, Uppsala, Sweden}
\affiliation{Department of Physics, Beijing Institute of Technology, Haidian District, Beijing 100081, P. R. China}


\begin{abstract}
~ 
\noindent
It's been said that  {\it mathematics is biology's next microscope, 
only better; biology is mathematics' next physics, only better} \cite{cohen-2004}.  Here we aim
for something  even better. We try to combine mathematical physics and biology into
a picoscope of life. For this we merge techniques which have been introduced and
developed in  modern mathematical physics, largely by Ludvig Faddeev
to describe objects such as solitons and Higgs and  to explain phenomena such as 
anomalies in gauge fields. We propose a synthesis that
can help to resolve the protein folding problem, one of the 
most important conundrums in all of science. 
We apply the concept of gauge invariance to scrutinize the extrinsic geometry of 
strings in three dimensional space.  We evoke general principles of symmetry in combination with
Wilsonian  universality and derive an essentially unique  Landau-Ginzburg energy that
describes the dynamics of a generic string-like configuration in the far infrared.
We observe that the energy supports topological solitons, that pertain to an anomaly in
the manner how a string is framed around its inflection points. 
We explain how  the solitons operate as modular building blocks 
from which folded proteins are composed. We describe 
crystallographic protein  structures by multi-solitons with experimental precision,  
and investigate the non-equilibrium dynamics of proteins under varying temperature. 
We simulate the folding process of a protein 
at {\it in vivo} speed and with close to pico-scale accuracy using a standard laptop 
computer: With  pico-biology as  mathematical physics' next pursuit, things can only get 
better.

\vskip 2cm
\noindent
This article is dedicated to Ludvig Faddeev on the occasion of his 80$^{th}$ birthday.
\end{abstract}

\maketitle

\section{Introduction}  
\label{sect1}

Ludvig Faddeev is the first to draw my attention to the following sentence in the
introduction of Dirac's stunning 1931 article \cite{dirac-1931}, 

\vskip 0.1cm
\begin{quote}
\noindent
"{\it There are at present fundamental problems in theoretical 
physics awaiting solution, e.g. the relativistic formulation of quantum mechanics and the nature of 
atomic nuclei (to be followed by more difficult ones such as the problem of life).}" 
\end{quote}
\vskip 0.1cm

\noindent
This  epitomizes what could be proclaimed as {\it the three Dirac problems 
in theoretical physics}.  The first Dirac problem was {\it de facto} solved at the time of writing.
Dirac introduced his equation in 1928. Maybe he already
had quantum gravity in his mind; it has been stated in exalted circles that this is a problem 
settled by string theory.
A resolution to the second Dirac problem, {\it the nature of atomic nuclei},  took almost half-a-century 
of collective efforts to develop. We now trust that LHC experiments at CERN demonstrate how
all four known fundamental interactions are governed by the Standard Model of Weinberg-Salam
and quantum chromodynamics, plus Einstein's gravity. However, there remain important conundrums 
awaiting solution, including that of quark (color) confinement.

The third Dirac problem, {\it  life}, endures. Its inclusion 
demonstrates Dirac's very high level of ambition, and trust 
on the mastery of theoretical physics. 
How could Dirac envisage  that the notion of  life could be defined as a theoretical physics problem? 
Did he presume that eventually life can be described with a conceptual clarity that
compares  with quantum mechanics?

Today we are reaching a point where a precise definition of {\it the problem of life}  could be attempted: We 
understand that proteins are the workhorses of all living cells. They 
participate in all the metabolic activities that constitute life, as we know it. 
We have learned that the biological function of a protein relates intimately to its shape. 
Thus  one might argue that  the {\it protein folding problem} is the way to
address the {\it problem of life} \'{a} la Dirac. 
The quest is to describe  the folding and dynamics of proteins
with the eloquence of equations in Dirac's 1931 article.

\section{Abelian Higgs Model}  
\label{sect2}
%

The Abelian Higgs Model (AHM) and its generalizations \cite{faddeev-1980}
comprise the paradigm 
theoretical framework for describing physical scenarios, from cosmic strings to 
vortices in superconductors. The Weinberg-Salam 
model of electroweak interactions is a non-Abelian epitome of AHM, and
so are the various  editions of Grand Unified Theory that aim to unify all sub-atomic 
forces. In the sequel we shall argue that a discretized version of AHM might even 
describe the dynamics and folding of proteins, 
with sub-atomic precision and at the {\it speed of life}.

The elemental AHM comprises 
a single complex scalar field $\phi$ and a vector field  $A_i$. 
These fields are
subjected to the U(1) gauge transformation
\begin{equation}
\begin{array}{lcl}
\phi \  & \to  & \ e^ {ie \, \vartheta } \phi\\
A_i \ & \to & A_i + \partial_i \vartheta
\end{array}
\label{u1}
\end{equation}
where $\vartheta$ is  a function, and $e$ is a parameter.
We 
may also introduce another set of variables $(J_i, \rho, \theta)$ that
relate to ($A_i,\phi$) by the following change of variables,
\begin{equation}
\begin{array}{lcl}
\phi \  & = & \ \rho \cdot e^ {i \theta } \\
A_i \ & \to & J_i = \frac{i}{2 e
|\phi|^2}\left[ \phi^* ( \partial_i -  i e A_i ) \phi - c.c. \right]
\end{array}
\label{asu}
\end{equation}
These new variables can be introduced whenever $\rho \not= 0$. The  
$\rho$ and $J_i$ are both gauge invariant,  they remain intact under 
the transformation (\ref{u1}). But
\[
\theta \ \to \ \theta + \vartheta
\]
The standard AHM Hamiltonian is
\begin{equation}
{\mathcal H} = \frac{1}{4} F^2_{ij} + |(\partial_i - i e A_i)
\phi|^2 + \lambda \left(|\phi|^2 -
v^2
\right)^2
\label{H1}
\end{equation}
where
\[
F_{ij} = \partial_i A_j - \partial_j A_i
\]
This is the  {\it unique} Landau-Ginsburg Hamiltonian
of the AHM multiplet, within the framework of Wilsonian universality \cite{kadanoff-1966,wilson-1971} and
invariance  under the U(1) gauge transformation (\ref{u1}), except that in 
odd dimensions $D$ a Chern-Simons term ($ChS$) could be added 
to break parity 
\begin{equation}
\begin{matrix} D = 1: & \ \ \ ChS & \sim & A & \\
D=3: & \ \ \ ChS & \sim &  AdA & \\
D=5: & \ \ \  ChS & \sim & AdAdA & \\ 
& & & & etc.
\end{matrix}
\label{chs}
\end{equation}

The gauge invariance of  (\ref{H1}) becomes manifest when we write it in terms of 
the new variables (\ref{asu}),
\begin{equation}
{\mathcal H} = \frac{1}{4} \left(
J_{ij} + \frac{2\pi}{e} \sigma_{ij}
\right)^2
\!\!
+ (\partial_i \rho)^2 + \rho^2 J_i^2
+ \lambda \left(\rho^2 - \eta^2 \right)^2 \ + \ ChS
\label{H2}
\end{equation}
Here
\[
J_{ij} = \partial_i J_j - \partial_j J_i
\]
and 
\be
\sigma_{ij} =
\frac{1}{2\pi}\, [\partial_i , \partial_j ] \theta
\lbl{ds1}
\ee
The $\sigma_{ij}$ is a string current, its support 
coincides with the world-sheet of the cores of 
Abrikosov vortices. 
When (\ref{H2}) describes  a vortex, (\ref{ds1}) subtracts a singular
string contribution that appears in $J_{ij}$. An irregular contribution then
appears in the third term in the {\it r.h.s.} of (\ref{H2}) but 
becomes removed by the density $\rho$ which
vanishes on the world-sheet of the vortex core.
Note that except along a vortex line the Hamiltonian (\ref{H2}) involves only variables
that are manifestly $U(1)$ gauge invariant. In particular, unlike in the case of
(\ref{H1}), in (\ref{H2}) the local gauge invariance is entirely removed. {\it Not} by 
fixing a gauge but by changing the variables \cite{chernodub-2008}.

\section{Strings and the Frenet Equation}
\label{sect3}

%

%
Proteins are string-like objects. Thus, to describe their physical properties we need to develop
the formalism of strings. We start with the continuum (differentiable) case.

The
geometry of a
class $\mathcal C^3$ differentiable string
$\mathbf x(z)$  in $\mathbb R^3$ is governed by the
Frenet equation \cite{frenet-1852}. 
Here $z \in [0,L]$ and  $L$ is the length of the string 
in $\mathbb R^3$ that is computed  by
\begin{equation}
L \ = \ \int\limits_0^L \! dz \,  || {\mathbf x}_z  ||  
\ = \  \int\limits_0^L \! dz \,  \sqrt{ {\mathbf x}_z \cdot {\mathbf x}_z } 
\ \equiv \ \int\limits_0^L \! dz \,  \sqrt{ g }.
\label{njl}
\end{equation} 
We recognize in (\ref{njl}) the (static) Nambu-Goto action;
the parameter $z \in [0,L]$ is generic. We may always
reparametrize the string and express it in terms of 
the arc-length $s \in [0,L]$ in the ambient
$\mathbb R^3$. This is achieved by the change of variables
\[
s(z) = \int\limits_0^z || {\mathbf x}_z (z') || d z'
\]
It the following we use the arc-length parametrization.
We shall also consider a single
open string and we assume that the string
does not self-cross {\it i.e.} it is self-avoiding.

At a generic point along the string, we introduce the unit length tangent vector
\begin{equation}
\mathbf t \ = \  
 \frac{ d \hskip 0.2mm \mathbf x (s)} {ds}  \ \equiv \ {\mathbf x}_s
\label{curve}
\end{equation}
the  unit length bi-normal vector
\begin{equation}
\mathbf b \ = \ \frac{  {\mathbf x}_s\times  {\mathbf x}_{ss} } { || 
 {\mathbf x}_s \times {\mathbf x}_{ss} || }
\label{bcont}
\end{equation}
and the unit length normal vector,
\begin{equation}
\mathbf n = \mathbf b \times \mathbf t
\label{ncont}
\end{equation} 
These three  vectors $(\mathbf n, \mathbf b, \mathbf t)$ form the right-handed orthonormal Frenet frame. 
This framing can be introduced at each point where
\begin{equation}
{\mathbf x}_s\times  {\mathbf x}_{ss} \ \not= \ 0
\label{kcond}
\end{equation}
We proceed, momentarily, by
assuming this to be the case.
The Frenet equation  \cite{frenet-1852} transepts  the frames along the string,
\begin{equation}
\frac{d}{ds}\!\left(
\begin{matrix} 
{\bf n} \\
{\bf b} \\
{\bf t} \end{matrix} \right) \ =  \   \left( \begin{matrix}
0 & \tau & - \kappa  \\ -\tau & 0 & 0 \\ \kappa & 0 & 0 \end{matrix} \right) 
\left(
\begin{matrix} 
{\bf n} \\
{\bf b} \\
{\bf t} \end{matrix} \right) 
\label{DS1}
\end{equation}
where 
\begin{equation}
\kappa(s) \ = \ \frac{ || {\mathbf x}_s \times {\mathbf x}_{ss} || } { ||  {\mathbf x}_s||^3 }
\label{kg}
\end{equation}
is the curvature of the string on the osculating plane that is spanned by $\mathbf t$ and $\mathbf n$, and
\begin{equation}
\tau(s) \ = \ \frac{ ( {\mathbf x}_s \times  {\mathbf  x}_{ss} ) \cdot { {\mathbf x}_{sss} }} { || {\mathbf x}_s 
\times  {\mathbf x}_{ss} ||^2 }
\label{tau}
\end{equation}
is the torsion that measures how the string deviates from its osculating plane. Both $\kappa(s)$ and 
$\tau(s)$  are  preordained solely by the extrinsic geometry {\it i.e.}  shape of the string. 
According to (\ref{DS1}), the curvature and torsion also
determine the 
string: If the two are known  we construct  $\mathbf t(s)$ by solving the Frenet equation
and then solve for
the string by integrating (\ref{curve}). The solution is unique up to rigid translations and 
rotations of the string.  

We note that the curvature (\ref{kg}) and the torsion (\ref{tau}) are
scalars under reparametrizations. Under an infinitesimal local diffeomorphism
along the string, obtained by deforming $s$ according to
\begin{equation}
s \to   s + \epsilon(s) 
\label{infi}
\end{equation}
where $\epsilon(s)$ is an arbitrary infinitesimally small function  such that
\[
\epsilon(0) = \epsilon (L) = 0 = \epsilon_s (0)  = \epsilon_s(L)
\]
we have 
\begin{equation}
\begin{matrix}
\delta \kappa (s) & = &  - \epsilon(s) \,  {\kappa}_s
\\ ~ \\ 
\delta \tau (s) & = & - \epsilon(s) \,  \tau_s \end{matrix}
\label{diffmor}
\end{equation}
The  Lie algebra of diffeomorphisms (\ref{infi})  is the 
classical Virasoro (Witt) algebra.

\section{Abelian gauge fields}
\label{sect4}

\begin{figure}[!hbtp]
  \begin{center}
    \resizebox{12.5cm}{!}{\includegraphics[]{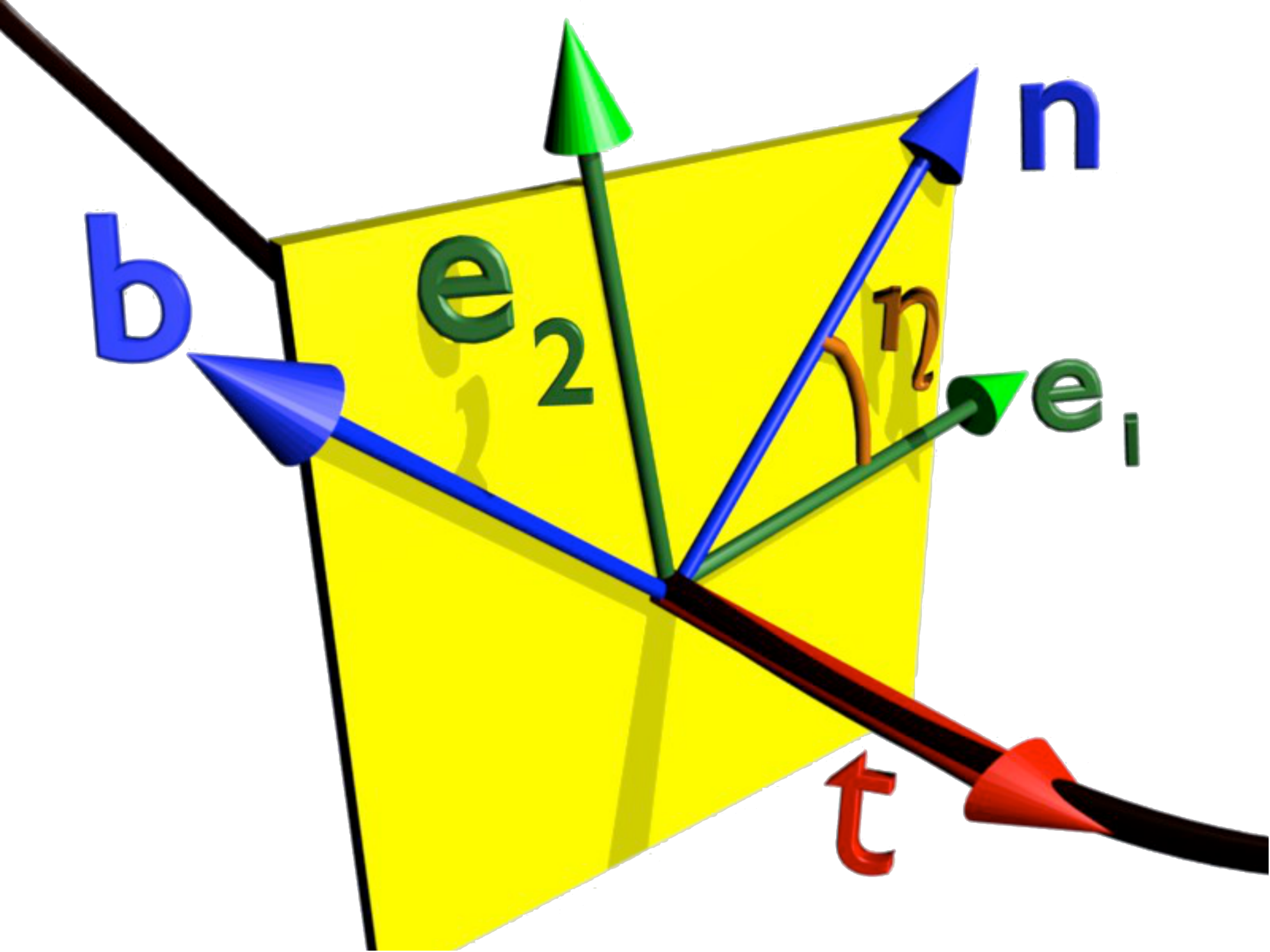}}
    \caption{Rotation between the Frenet frames and generic frames on the normal plane of the string.} 
    \label{fig1}
  \end{center}
\end{figure}

From the extrinsic string geometry we can construct an exemplar Abelian Higgs Multiplet
\cite{niemi-2003,hu-2013,ioannidou-2014}. 
For this we observe that (\ref{curve}) 
does not engage the normal and bi-normal vectors. Thus a SO(2) 
rotation around $\mathbf t (s)$ that sends the 
zweibein ($\mathbf n , \mathbf b$) into 
another zweibein ($\mathbf e_1 , \mathbf e_2 $)  as shown in figure \ref{fig1}, 
leaves the string intact.
%
%
%

 The  generic zweibein is  obtained from the Frenet zweibein as follows,
\begin{equation}
\left( \begin{matrix} {\bf n} \\ {\bf b} \end{matrix} \right) \ \to \ \left( \begin{matrix} {{\bf 
e}_1} \\ {\bf e}_2 \end{matrix} \right) \
= \ \left( \begin{matrix} \cos \eta(s) &  \sin \eta(s) \\  - \sin \eta(s) & \cos \eta(s) \end{matrix}\right)
\left( \begin{matrix} {\bf n} \\ {\bf b} \end{matrix} \right).
\label{newframe}
\end{equation}
For  the Frenet equation this yields
\begin{equation}
\frac{d}{ds} \left( \begin{matrix} {\bf e}_1 \\ {\bf e }_2 \\ {\bf t} \end{matrix}
\right) =
\left( \begin{matrix} 0 & (\tau + \eta_s) & - \kappa \cos \eta \\ 
- (\tau + \eta_s)  & 0 & \kappa \sin \eta \\
\kappa \cos \eta &  - \kappa \sin \eta  & 0 \end{matrix} \right)  
\left( \begin{matrix} {\bf e}_1  \\ {\bf e }_2 \\ {\bf t} \end{matrix}
\right).
\label{contso2}
\end{equation}
Remarkably, by interpreting $\kappa(s)$ as the modulus of a complex quantity
\begin{equation}
\kappa \ \buildrel{\eta}\over\longrightarrow \ \kappa e^{-i\eta}
\label{keta}
\end{equation}
we may consider the transition of ($\kappa,\tau$) in (\ref{newframe}) as 
a one-dimensional example of the U(1) gauge
transformation (\ref{u1}), with curvature identified as the Higgs field  and torsion as the gauge field,
\begin{equation}
\begin{matrix} \kappa & \sim & \phi  &  \to & \kappa e^{-i\eta} & \equiv  &  \phi \, e^{-i\eta} \\ ~ \\
\tau & \sim &  A_i &  \to & \tau + \eta_s & \equiv  &   A_i + \eta_s
\end{matrix}
\label{ahmulti}
\end{equation}
We remark that the choice  
\begin{equation}
\eta(s) =  - \int_0^s \! \tau (\tilde s) d\tilde s 
\label{partran}
\end{equation}
brings about  the {\it unitary}   gauge, that
corresponds to the parallel transport framing \cite{bishop-1974}. Unlike the Frenet 
framing that can not be defined at points (or segments) where (\ref{kcond}) vanishes,
the parallel transport framing can still be defined in a continuous manner  
\cite{hanson}.  But there is an anomaly lurking, that we soon reveal.

\section{Spinor Frenet Equation}
\label{sect5}

Occasionally, it is profitable to recognize that the 
Frenet equation admits a spinor representation \cite{ioannidou-2014}. 
For this we introduce a two-component 
complex spinor 
\begin{equation}
\psi =  \left( \begin{matrix} z_1  \\ z_2 \end{matrix}\right) 
\label{spinor}
\end{equation}
We also introduce the conjugate spinor $\bar\psi$; the two are related by 
the charge conjugation operation $\mathcal C$,
\begin{equation}
\bar\psi \ = \ \mathcal C \psi \ = \ -i \sigma^2 \psi^\star 
\ = \ \left( \begin{matrix} -  z_2^\star  \\ z_1^\star \end{matrix}\right) 
\label{spinor2}
\end{equation}
Note that
\[
\mathcal C^2 = - \mathbb I
\]
We impose  the normalization condition
\begin{equation}
\psi^\dagger \psi  \ \equiv \ <\psi,\psi> = 1 = <\bar\psi , \bar\psi> = \bar\psi^\dagger \bar\psi
\label{ortho}
\end{equation}
\[
<\psi \, , \bar\psi> = 0
\]
We define the spin polarization vector $ {\mathbf t}$ 
so that
\begin{equation}
{\mathbf t} \cdot \hat \sigma \, \psi = \psi \ \ \ \ \ \Leftrightarrow \ \ \ \ \
{\mathbf t} \cdot \hat \sigma \, \bar\psi =  - \bar\psi
\label{ts}
\end{equation}
We get
\[
{\mathbf t} = \psi^\dagger \hat \sigma \psi  \ = \ <\psi, \hat\sigma \psi>
\]
We also define the following complex vector,
\[
{\mathbf e}_+  \ = \ {\mathbf e}_1 + i {\mathbf e}_2 \ = \ <\bar \psi , \hat \sigma \psi >   
\ \equiv \ {\bar \psi}^\dagger \hat\sigma \psi
\]
where $\mathbf e_1$ and $\mathbf e_2$ are real.  We can check that 
\[ 
{\mathbf e}_i \cdot {\mathbf e}_j = \delta_{ij} \ \ \ \ \ \& \ \ \ \ \ {\mathbf e}_i \cdot {\mathbf t} = 0
\]
and we conclude that ($\mathbf e_1, \mathbf e_2, \mathbf t$) is a right-handed orthonormal system.
It can be identified with a Frenet framing of a string defined by $\mathbf t(s)$ as the tangent vector,
possibly modulo a global SO(2) frame rotation.

\vskip 0.2cm
Consider the following  local U(1) rotation:
\begin{equation}
\psi \to e^{i\eta} \psi \ \ \ \ \& \ \ \ \ \bar\psi \to e^{-i\eta} \bar\psi
\label{u11}
\end{equation}
Then,
\[
<\psi, \partial_s \psi> = \psi^\dagger \partial_s \psi \ \to \ <\psi, \partial_s \psi>  + i \partial_s \eta
\]
This suggest we introduce  the putative \emph{torsion}
\begin{equation}
\tau \sim - i <\psi, \partial_s \psi> 
\label{puttor}
\end{equation}
Furthermore, since
\[
<\bar\psi , \partial_s \psi > = \bar\psi^\dagger \partial_s \psi \ \to \ 
e^{2i\theta} <\bar\psi , \partial_s \psi > 
\]
we identify 
\begin{equation}
\kappa \ \sim \ <\bar\psi , \partial_s \psi > \ \sim \ \kappa
\label{putcur}
\end{equation}
as the putative (complex) \emph{curvature}.  
Thus, we have the \emph{spinor Frenet equation}: 
\begin{equation}
\partial_s \psi \ = \  i \tau  \psi + \kappa \bar \psi 
\label{frenet1}
\end{equation}
If curvature and torsion in (\ref{frenet1}) are known, we can solve for $\psi$ and compute 
\[
{\mathbf t} = <\psi, \hat \sigma \psi>
\]
The string in the arc-length parametrization is then determined as before, from
\[
 \frac{ d \mathbf x}{ds} = {\mathbf t} 
\]

\section{Non-abelian gauge fields}

\label{sect6}

The geometric interpretation of curvature and torsion  in terms of the AHM multiplet
extends to a non-Abelian framework. This brings about a relation between 
extrinsic string geometry and the structure of non-Abelian gauge theories, in a decomposed 
format of the latter \cite{faddeev-1998,faddeev-1999-a,faddeev-1999-b,faddeev-2007} . 
These relations are valuable for a wider 
perspective, beyond the immediate scope of the present article:
We start by introducing the three  matrices 
\begin{equation}
T_1 = \left( \begin{matrix} 0 & 0 & 0 \\
0 & 0 & -1 \\ 0 & 1 & 0 \end{matrix} \right), \ \  T_2 = \left( \begin{matrix} 0 & 0 & 1 \\
0 & 0 & 0 \\ -1 & 0 & 0 \end{matrix} \right), \ \  T_3 = \left( \begin{matrix} 0 & -1 & 0 \\
1 & 0 & 0 \\ 0 & 0 & 0 \end{matrix} \right)
\label{T}
\end{equation}
that determine the canonical adjoint representation of \underline{so}(3) Lie algebra,
\[
\left[T_a , T_b \right] = \epsilon_{abc} T_c.
\]
The action of the SO(2)$\sim$U(1) frame rotation on  $\kappa$ and $\tau$  may be 
realized as follows,
\begin{eqnarray}
\hspace{-18mm} \kappa T_2  & \to &
e^{-\eta T_3} \left( \kappa\, T_2 \right)  e^{\eta T_3}  =  \kappa \left( \cos \eta\, T^2 - \sin \eta\, T^1\right),
\label{sokna}\\
\hspace{-8mm}\tau \, T_3 & \to & \left(\tau + \eta^\prime \right)  T_3.
\label{sotna}
\end{eqnarray}
This proposes that we combine the (general frame) curvature and torsion into a
non-Abelian SO(3) gauge field
\begin{equation}
A^a T^a \ = \ A^1 T^1 + A^2 T^2 + A^3 T^3  \ = \ \kappa \, \sin\eta \, T^1 +  \kappa \, \cos\eta \, T^2 + \tau_\eta T^3
\label{ASO3}
\end{equation}
The SO(3) gauge transformation 
\[
A^aT^a \ \to \ g^{-1} ( A^a T^a + i  \partial_s ) g
\]
corresponds to a general frame rotation while (\ref{sokna}), (\ref{sotna})  
determine the subset of SO(2) gauge transformations in the
Cartan direction
\[
g \ = \ e^{\eta T^3}
\]

The spinor Frenet equation can be interpreted as follows:  We combine the two spinors (\ref{spinor}), (\ref{spinor2})
into a four component Majorana spinor 
\[
\Psi = \frac{1}{\sqrt{2}} \left( \begin{matrix} \psi \\ \bar\psi \end{matrix} \right) 
\]
Using (\ref{puttor}), (\ref{putcur}) we recover the connection (\ref{ASO3}),  now in the SU(2) basis,
\[
A^a T^a \ \sim \ \mathcal A_{\alpha\beta} = \Psi^\dagger_\alpha \partial_s \Psi_\beta \ = \ \left( \begin{matrix}  \tau & - i 
\kappa \\ 
i \kappa^\star & - \tau 
\end{matrix} \right) \ = \ {\mathcal A} \cdot \hat \sigma 
\]
\begin{equation}
 = \ \tau \sigma^3 + \kappa_1 \sigma^1 + \kappa_2 
\sigma^2
 \equiv \ \tau \sigma^3 + \kappa \sigma^+ + \kappa^\star \sigma^-
\label{conn}
\end{equation}

\[ 
\kappa_1 = \Re e [\kappa]  \ \ \ \ \  \&  \ \ \ \ \ \kappa_2 \ = \ \Im m [\kappa]
\]

\noindent
Then we have from (\ref{frenet1}) for the Frenet equation
\begin{equation}
\left( i \partial_s + {\mathcal A} \cdot \hat\sigma \right) \Psi = 0
\label{ws}
\end{equation}
which remains covariant under the  SU(2) gauge transformation
\[
\Psi \to g \Psi \ \equiv \Psi_g \ \ \ \ \ \Rightarrow \ \ \ \ \ \mathcal A \ \to \ g ( \mathcal A  + i \partial_s ) g^{-1} \ \equiv \ \mathcal A_g
\]
corresponding to a general SO(3)$\simeq$SU(2) rotation of frames, along the string. 
Of particular interest is the gauge transformation
defined by $g \in$ SU(2) so that
\[
g \, \sigma^3 g^{-1} \ = \ \mathbf t \cdot \hat \sigma \ \equiv \hat{\mathbf t} 
\]
\[
g \, ( \sigma^1 + i \sigma^2) g^{-1} \ = \ (\mathbf n + i \mathbf b) \cdot \hat \sigma \ \equiv 
\ {\mathbf e}_+ \cdot \hat \sigma \ = \ \hat{\mathbf e}_+ 
\]
This identifies the Frenet framing and sends (\ref{conn}) into
\begin{equation}
{\mathcal A} \cdot \hat \sigma \ \to \ ( \tau \, \hat{\mathbf t} + \kappa \, \hat{\mathbf e}_+ + \kappa^\star
\hat{\mathbf e}_-)  + {\mathfrak a}\, \hat{\mathbf t} + \frac{1}{2i} [ \partial_s \hat{\mathbf t} , \hat{\mathbf t}]
\label{fniemi}
\end{equation}
\[
{\mathfrak a} \ = \ 2i < \mathbf e^+ , \partial_s \mathbf e^->
\]

\noindent
The SU(2) connection (\ref{fniemi}) has the functional form of the
decomposed connection introduced in \cite{faddeev-1998}, including the "monopole" contribution.  In particular, 
the U(1)$\in$ SU(2) gauge rotation around the Cartan direction $\hat{\mathbf t}$ \cite{faddeev-1998}
\[
g \ = \ \exp\{ i \frac{\eta}{2}  \, \hat{\mathbf t} \}
\]
sends
\[
\begin{matrix} \kappa &  \to &  e^{i\eta} \kappa \\
\tau & \to & \tau + \partial_s \eta \\
\mathfrak a & \to & \mathfrak a + \partial_s \eta
\end{matrix}
 \]
This is the same as (\ref{newframe}), (\ref{contso2}).  The present construction divulges that there 
is an intrinsic string design hiding in non-Abelian gauge theories. 

\section{energy of string}
\label{sect7}

To describe string dynamics, and eventually that of proteins, we desire an energy function in terms of
the curvature and torsion. For this we remind that the Landau-Ginzburg 
energy  (\ref{H1})  of the elemental AHM  is {\it unique} in the Wilsonian
sense of universality. It emerges from general arguments and 
symmetry principles alone.  We  also remind
that the shape of a string does not depend on the way 
how we  frame it. 
Accordingly,  the energy of a string can only engage manifestly frame independent  
combinations of the curvature and torsion (\ref{ahmulti}). 
Since (\ref{H1}) is the 
universal SO(2)$\sim$U(1) invariant energy function that involves a complex  Higgs field
$\phi \sim \kappa$ and a gauge field $A\sim  \tau$, it must serve as the Hamiltonian that
describes strings and their dynamics in the far infrared. Thus, we introduce
\begin{equation}
H \ = \ \int\limits_0^L ds \, \left \{ \, |(\partial_s + i e \,{ \tau})  { \kappa} |^2 + \lambda\, (|
{\kappa} |^2 - m^2)^2 \, \right \}
\ + \ a \! \int\limits_0^L ds \, { \tau}
\label{enes}
\end{equation}
We have  included  the one dimensional version of  the Chern-Simons term (\ref{chs}), it introduces chirality to the
string;  in one dimension there is no $F_{ij}$. 

In (\ref{enes}), the curvature $\kappa$ and torsion $\tau$ are expressed 
in a generic, arbitrary framing of the string.
The ensuing gauge invariant variables (\ref{asu})
coincide with the Frenet frame curvature (\ref{kg}) and torsion (\ref{tau}) that characterize the
extrinsic string geometry. In terms of these {\it gauge invariant variables}, which we  {\it from now on} denote
by ($\kappa,\tau$)  
the Hamiltonian (\ref{enes}) becomes 
\begin{equation}
H \ = \ \int\limits_0^L ds \, \left \{ \, (\partial_s { \kappa})^2  + e^2 {\kappa}^2 {\tau}^2 + 
\lambda\, ({\kappa}^2 - m^2)^2 \, \right \}
\ + \ a \int\limits_0^L ds \, \tau
\label{enes2}
\end{equation}
This is our manifestly gauge invariant Landau-Ginzburg Hamiltonian  of strings; see (\ref{H2}).

Finally, we point out  a  non-Abelian variant of the energy:  
A solution to the equation (\ref{ws}) minimizes the functional
\[
 \mathcal S  = \int ds \ | (i\partial_s + \mathcal A) \Psi |^2
\]
This is an exemplar of the gauged SU(2) non-linear $\sigma$-model; recall that the Majorana spinor
is subject to the normalization condition  that derives from (\ref{ortho}). 
Additional SU(2) gauge invariant functionals of ($\Psi ,  \mathcal A$) could 
be introduced. This   leads  to 
gauged non-linear $\sigma$-models as energy functions of strings.

\section{Integrable hierarchy}
\label{sect8}

There is a relation between (\ref{enes2}),  the integrable hierarchy of the 
nonlinear Schr\"odinger (NLS) equation, and the Heisenberg spin chain. For this we follow Hasimoto 
\cite{hasimoto-1972} and combine the curvature and torsion into the complex variable
\begin{equation}
\psi(s) \ = \ \kappa(s)\,  e^{i e\int\limits_0^s \, ds' \, \tau(s') }
\label{hasi}
\end{equation} 
This yields us the standard NLS Hamiltonian density \cite{faddeev-1987,ablowitz-2003}
\begin{equation}
\kappa_s^2 + e^2 \kappa^2 \tau^2  + \lambda \kappa^4 \ = \  \bar\psi_s \psi_s + \lambda (\bar\psi \psi)^2 \ = \ 
\mathcal H_3 
\label{H3}
\end{equation}
The lower level conserved densities in the integrable NLS hierarchy 
are the helicity, length ({\it i.e.} Nambu-Goto), 
number operator and momentum respectively
\begin{equation}
\begin{matrix} 
\mathcal H_{-2} & = & \tau \\
\mathcal H_{-1} & = & L \\
\mathcal H_{1} & = & \kappa^2 \\
\mathcal H_{2} & = & \kappa^2 \tau
\end{matrix}
\label{Hnls}
\end{equation}
The energy (\ref{enes2}) is a combination of $\mathcal H_{-2}$, $\mathcal H_{1}$ and $\mathcal H_{3}$. From the
perspective of the NLS hierarchy, 
the momentum $\mathcal H_{2}$ could also be included. In case higher order corrections are
desired, the natural candidate is the mKdV density
\[
\mathcal H_4 \ = \ \kappa \kappa_{ss} \tau + 4 \kappa^2 \tau^3 - 4e^2 \kappa_s^2 \tau + 3 \lambda \kappa^4 \tau
\]
with appears as the next level conserved density in the NLS hierarchy. 

The Heisenberg spin chain is obtained
directly from $\mathcal H_{1}$, in terms of the Frenet equation.
\[
\int\limits_0^L ds \, \mathcal H_{1}  \  =   \int\limits_0^L ds \, \kappa^2 \ = \  \int\limits_0^L ds \, |\mathbf t_s|^2
\]
The combination  of $\mathcal H_{-1}$ and ${\mathcal H}_1$ leads to the rigid string action \cite{polyakov-1986}, 
it also relates to Kratky-Porod model of polymer physics \cite{kratky-1949}. In \cite{polyakov-1986},
perturbative level Wilsonian universality is employed to argue that no additional terms should emerge in the infrared. 
However, a perturbative approach does not reach into the non-perturbative, the 
NLS Hamiltonian  (\ref{H3}) is known to support solitons that are pivotal in the description 
of numerous far infrared physical phenomena.

\section{Solitons}
\label{sect9}

Solitons are the paradigm structural
self-organizers in the physical world. They materialize in numerous scenarios 
\cite{faddeev-1987,ablowitz-2003,manton-2004,kevrekidis-2009}:
Solitons transmit data in transoceanic cables, solitons conduct 
electricity in organic polymers and transport chemical energy in proteins.
Solitons explain the Mei\ss ner effect in  superconductivity and dislocations in liquid crystals. 
Solitons model hadronic particles, cosmic strings and magnetic monopoles in  high  energy physics.  
In the sequel we argue that  solitons might even describe life: 
There are about $10^{20}$ or so solitons in your own  human body. They participate 
in all the metabolic processes that make you kick, spark and to be alive.
 
The NLS equation that derives from (\ref{H3}) is the paradigm equation that 
supports solitons \cite{faddeev-1987,ablowitz-2003,manton-2004,kevrekidis-2009}.
Depending on the sign of $\lambda$, these solitons 
are either dark ($\lambda >0$) or bright ($\lambda <0$).
Furthermore, on the real line 
$\mathbb R$ the torsion independent contribution to (\ref{enes2})
\begin{equation}
H \ = \ \int\limits_{-\infty}^\infty ds \, \left \{ \,  \kappa_s^2   + 
\lambda\, (\kappa^2 - m^2)^2 \, \right \}
\label{swave}
\end{equation}
describes the double well kink {\it a.k.a.}  the paradigm {\it topological} soliton:
For positive $m^2$ and when $\kappa$ can take both positive and negative values 
the ensuing equation of motion 
\[
\kappa_{ss} =  2 \lambda \kappa (\kappa^2 - m^2)
\]
is solved by 
\begin{equation}
\kappa(s) \ = \ m \, \tanh \left[ m \sqrt{\lambda} (s-s_0)\right]
\label{kink}
\end{equation}

The energy function
\begin{equation}
\mathcal E \ = \ \int ds \, \left\{ \, \kappa_s^2 + \lambda (\kappa^2 - m^2)^2 + \frac{d}{2} \kappa^2 \tau^2
- b \kappa^2 \tau - a\tau + \frac{c}{2} \tau^2 \, \right\}
\label{enenls}
\end{equation}
is the most general linear combination of {\it all} the densities (\ref{H3}), (\ref{Hnls}). 
In addition we have included 
the Proca mass term (the last term) \cite{hu-2013,ioannidou-2014}.  
Even though the Proca mass does 
not appear in the integrable NLS hierarchy, it does 
have a claim of gauge invariance and as such it is often accounted for. 
Since (\ref{enenls}) is quadratic in the torsion, we may eliminate 
$\tau$ using the ensuing  equation of motion,
\begin{equation}
\frac{\delta \mathcal E}{\delta \tau} \ = \ d \kappa^2 \tau - b \kappa^2 - a + c\tau \ = \ 0
\label{moneq0}
\end{equation}
This  gives
\begin{equation}
\tau [\kappa] \ = \ \frac{ a + b\kappa^2}{c+d\kappa^2}
\label{taueq}
\end{equation}
Thus we obtain the following {\it effective} energy for the curvature,
\begin{equation}
\mathcal E_\kappa \ = \ \int ds \, \left\{ \frac{1}{2} \kappa_s^2 + V[\kappa]  \right\}
\label{modNLS}
\end{equation}
with the equation of motion
\begin{equation}
\frac{\delta \mathcal E_\kappa}{\delta \kappa} \ = \ -\kappa_{ss} +  V_\kappa \ = \ 0
\label{monequ}
\end{equation}
where
\begin{equation}
V[\kappa] \ = \ - \left( \frac{ bc - ad}{d}\right) \, \frac{1}{c+d\kappa^2} \ - \ \left( \frac{b^2 + 8\lambda m^2}{2b} \right)
\, \kappa^2 + \lambda \, \kappa^4
\label{V}
\end{equation}
This is a deformation of the potential in (\ref{swave});  the two share the same large-$\kappa$ asymptotics. 
For a suitable choice of parameters we expect that (\ref{moneq0}), (\ref{monequ}), (\ref{V}) continue to support topological solitons.
But we do not know their explicit profile, in terms of elementary functions. In the sequel we construct the solitons numerically.

\section{Frame anomaly}
\label{sect10}

We now  introduce the  frame anomaly. It is  the edifice of a topological 
soliton along the string. Thus far we have tacitly assumed (\ref{kcond}). It ensures
that the Frenet curvature (\ref{kg}) does not vanish. But  we already 
pointed out that in AHM we have
$\rho = 0$ at the core of a vortex line.  Moreover, the explicit profile  (\ref{kink}) 
is  both positive and negative, 
and vanishes at $s=s_0$. Thus the consequences of $\kappa = 0$ for a
string deserves to be investigated. For this it 
can be profitable to extend the Frenet curvature $\kappa(s)$ so that it takes 
both positive and negative values. From  (\ref{keta})
we conclude that the negative values of $\kappa$ are related to 
the positive ones by a 
$\eta =  \, \pm\pi $ frame rotation,
\begin{equation}
\kappa \ \buildrel{\eta\, = \, \pm \pi}\over\longrightarrow \ e^{\pm i \pi } \kappa \ = \ - \kappa 
\label{kappaeta}
\end{equation}
Hence we compensate for the
extension of $\kappa$ to negative values, by engaging a 
discrete $\mathbb Z_2$ symmetry \cite{hu-2011}.
 
\begin{figure}[!hbtp]
  \begin{center}
    \resizebox{12.5cm}{!}{\includegraphics[]{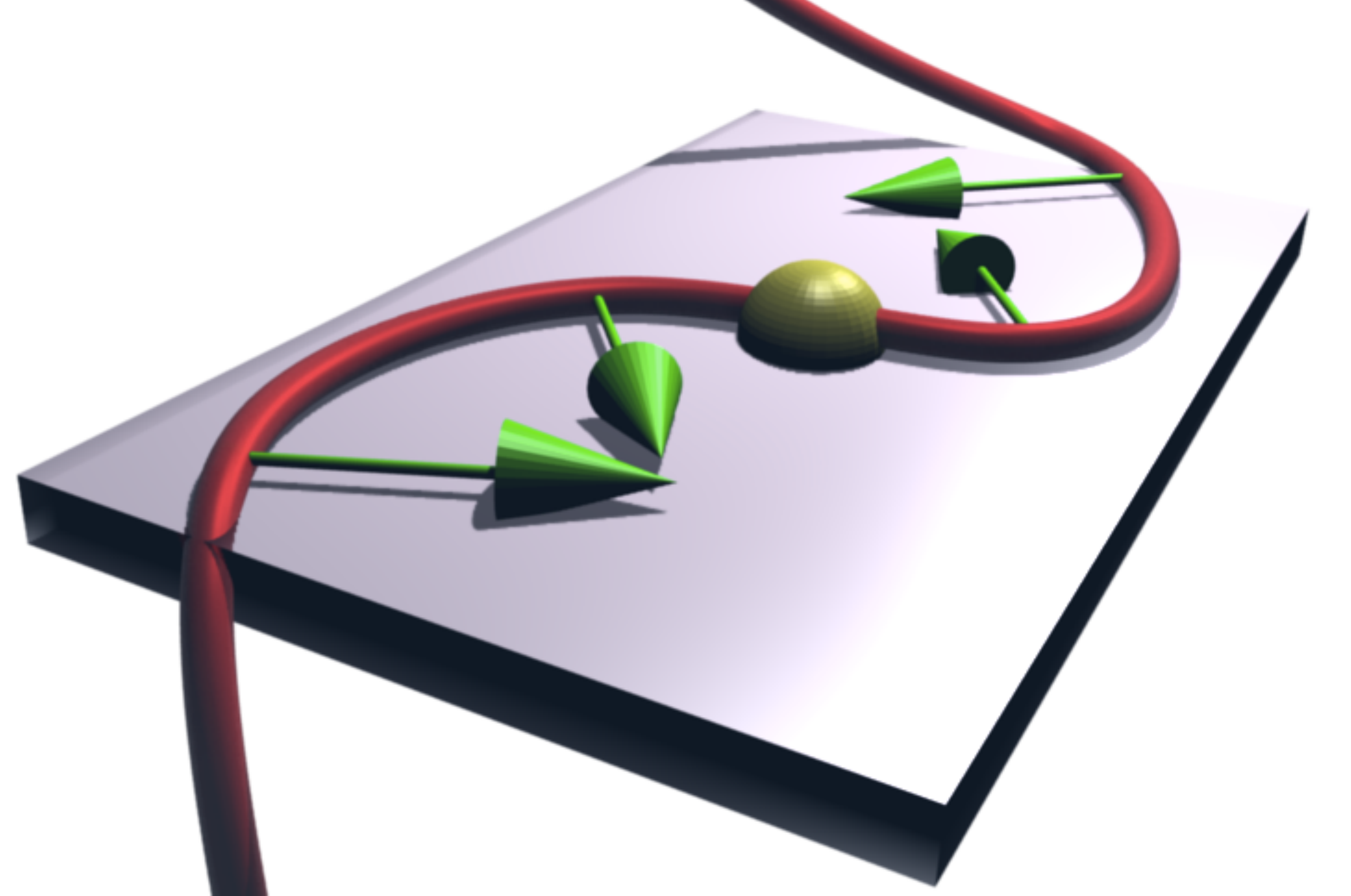}}
    \caption{A string with inflection point (ball). At each point the Frenet frame normal vector points towards the center of
    the osculating circle. At the inflection point there is a discontinuity in the direction of the normal vectors: 
    The radius of the osculating  circle diverges and the normal vector are
    reflected in the osculating plane, from one side to the other side of the string.      } 
    \label{fig2}
  \end{center}
\end{figure}

An (isolated) point where the curvature $\kappa(s)$ vanishes is 
an inflection point. In figure \ref{fig2}
we show an example of an inflection point.
Notice how the Frenet framing experiences a sudden change:  
The zweibein ($\mathbf n, \mathbf b$) is reflected according to
\begin{equation}
(\mathbf n + i \mathbf b)  \ 
\longrightarrow \  - (\mathbf n + i \mathbf b) \ = \  e^{\pm i \pi}(\mathbf n + i \mathbf b)
\label{anom}
\end{equation}
when it proceeds through the inflection point. At the inflection point itself the 
Frenet frame is  not defined. Thus  we can not directly determine whether 
there has been a jump by $\eta = +\pi$ or by $\eta = -\pi$ in figure \ref{fig2}. That is, does the frame turn left or does it turn
right by $\pi$, at the inflection point. We have a {\it frame anomaly}.

To scrutinize this anomaly, we  consider a  string $\mathbf x(s)$ that has 
a simple inflection point when $s=s_0$ so that $\kappa(s_0)=0$ but $\kappa_s(s_0) \not=0$;
for notational simplicity we here re-define the parameter $s$ so that $s_0=0$.  
Since a generic string in $\mathbb R^3$ has no inflection point, we may also
remove it from $\mathbf x(s)$  by a  
generic deformation. We introduce two such deformations,
\begin{equation}
\mathbf x(s) \ \to \ \mathbf x(s) + \delta \mathbf x_{1,2}(s) \ = \ \mathbf x_{1,2}(s)
\label{12curv}
\end{equation}
In figure \ref{fig2} these deformations would amount to a move of the string either 
slightly up from the plane or slightly down from the plane, to remove the inflection point; a move 
restricted to the plane only slides the inflection point.
We assume the deformations are tiny and compactly supported so that 
\[
\delta \mathbf x_{1,2}(\pm \varepsilon_\pm) = 0
\] 
Here  $\varepsilon_\pm>0$   
are small and determine the parameter values where the deformations 
$ \mathbf x_{1,2}(s)$ bifurcate. 
We consider a closed string $\Gamma$ that starts from 
$\mathbf x(-\varepsilon_- )$, follows along $\mathbf x_1 $ to $\mathbf x(+\varepsilon_+ )$
and then returns along $\mathbf x_2 $  back to 
$\mathbf x(-\varepsilon_-)$. We introduce the Frenet frame 
normal vectors of $\Gamma$, to define a 
second closed string $\tilde\Gamma$. It is obtained by shifting $\Gamma$ slightly 
into the direction of its Frenet frame normals. Let  $\mathbf t$, $\tilde{\mathbf t}$ be the ensuing 
tangent vectors.  We compute  
the Gau\ss~linking number 
\[
{\tt Lk} \ = \  \frac{1}{4\pi} \oint\limits_{\Gamma} \oint\limits_{\tilde \Gamma} ds d\tilde s \,
\frac{\mathbf x - \tilde{\mathbf x} }{|\mathbf x - \tilde{\mathbf x} |^3} \cdot (\mathbf t  \times \tilde {\mathbf t })
\]
When we proceed along $\mathbf x_{1,2}(s)$ the respective Frenet frames become continuously 
rotated by $\eta_{1,2}\approx \pm \pi$; in the limit 
where $\delta \mathbf x_{1,2} \to 0$ we obtain
the discontinuous jump (\ref{anom}). 
By continuity of Frenet framing in the complement 
of inflection points, the linking number has  values  {\tt Lk}$=\pm 1$ when the $\eta_{1,2}$ 
change in the same direction; recall that $\Gamma$ proceeds "backwards" 
along $\mathbf  x_2 $. But {\tt Lk}$=0$ if the framing along $\mathbf x_1(s)$ and $\mathbf x_2(s)$ rotate
in the opposite directions.  
Crucially the relative sign of $\eta_{1,2}$ appears to depend on the way 
how the inflection point becomes circumvented. Thus there is a frame anomaly as $\delta \mathbf x_{1,2}\to 0$.
The value of {\tt Lk} depends on the way  how we define $\delta \mathbf x_{1,2}(s)$.

\section{Perestroika's}
\label{sect10a}

When an inflection point occurs and a frame anomaly takes place,  we have a
string specific bifurcation which is called {\it inflection point perestroika} 
\cite{arnold1,arnold2,arnold3,aicardi,uribe}.
It renders futile our  attempts to uniquely  
frame the string $\mathbf x(s)$ across the inflection point. 
But there is also another kind of perestroika that takes place at the inflection point, which we now explain:

We start with a long flat string segment, so that the torsion $\tau(s)$ of the segment 
vanishes. This is synonymous for the segment to be constrained on a plane, {\it e.g.} as shown in figure 
\ref{fig2}.  For a string on the plane a simple isolated inflection is generically present, 
somewhere along the string.  Moreover, if we deform the string  but strictly  in a manner that retains
it on the plane, a simple inflection point can not disappear. 
It only moves around, unless it escapes thru the ends of the string which we 
assume is not the case:
For a string on the plane, a single inflection point  
is a topological invariant.  

Consider now the string in $\mathbb R^3$. Generically, it does not have any inflection points. But 
if the string moves freely, an isolated simple
inflection point generically appears at some isolated value of the flow parameter. 
Furthermore, when the ensuing infection point 
perestroika takes place along  the moving string, it leaves behind a trail: 
The {\it momentary} inflection point 
perestroika {\it permanently} changes the number of  {\it flattening points} which are 
the points along the string where its torsion $\tau(s)$ vanishes \cite{aicardi,uribe}. 

At a simple flattening point the curvature $\kappa(s)$ is generically non-vanishing, while the torsion $\tau(s)$ 
changes its sign. Thus  the inflection point perestroika can only change the   
number of simple flattening points by two. Apparently, it always does \cite{aicardi,uribe}.

Unlike the inflection point,
a flattening point is generic in a static space string.
Furthermore, unlike a simple inflection point, a single 
simple flattening point that occurs in a one parameter family of strings in $\mathbb R^3$
is a topological invariant. It can not disappear on its own, under local 
deformations that do not touch the ends of the string.
But a pair of flattening points may become combined into a single bi-flattening point which can 
dissolve. When this  happens, a second string-specific  bifurcation 
called {\it bi-flattening perestroika} takes place.

Apparently, inflection point perestroika and bi-flattening perestroika are the only two bifurcations 
where the number of flattening points can change \cite{uribe}.
The number of simple flattening 
points is a local invariant of the string. Besides the 
flattening number, and  the self-linking number in case of a framed string, a generic 
smooth string does not possess any other essential 
local invariants \cite{aicardi}. The two are also mutually independent, even though they often conspire.

For example, the self-linking number of the string increases by one if the torsion is
positive when the string approaches its simple inflection point, and if two simple 
flattening points disappear after the passage of the inflection point. 
Moreover, if the torsion is negative, the self-linking number decreases by one 
when two flattening points disappear after the 
passage \cite{aicardi}. But when two simple flattening points dissolve
in a bi-flattening perestroika, the self-linking number in general does not change. 
 
We conclude this Section with the following {\it statement}: 
Bifurcations are the paradigm causes for restructuring transitions in any
dynamical system.  Perestroika's are the only known 
stringy versions of bifurcations. Thus perestroika's have a potentially profound influence 
on the physical behavior of string-like structures. Moreover, perestroika's relate
to the frame anomaly which is a structural attribute of a string. This identifies  perestroika's  
as those bifurcations, that drive string-specific phase transitions which involve 
structural re-organizations.  In particular, perestroika's prompt topological solitons 
such as (\ref{kink}) to come and go along a string. Such 
processes are commonplace whenever we have an energy function of the 
form (\ref{enenls}). It appears that 
in the case of strings that model proteins, we always do.  In proteins, we find
solitons and perestroika's all over the place.

\section{Discrete Frenet Frames}
\label{sect11}

Proteins are linear macromolecules with a highly complex chemical composition. Proteins are
made of twenty different covalently bonded amino acid molecules (residues) that are  
joined together in a row, one after another. But despite the diversity of amino acids 
most proteins share a plenty of conformational similarities. To an extent, that their
structure and dynamics can be described using a single theoretical
model that concurs with a distinct universality class,  in the 
sense of Kadanoff and Wilson \cite{kadanoff-1966,wilson-1971}. 

\begin{figure}[!hbtp]
  \begin{center}
    \resizebox{12.5cm}{!}{\includegraphics[]{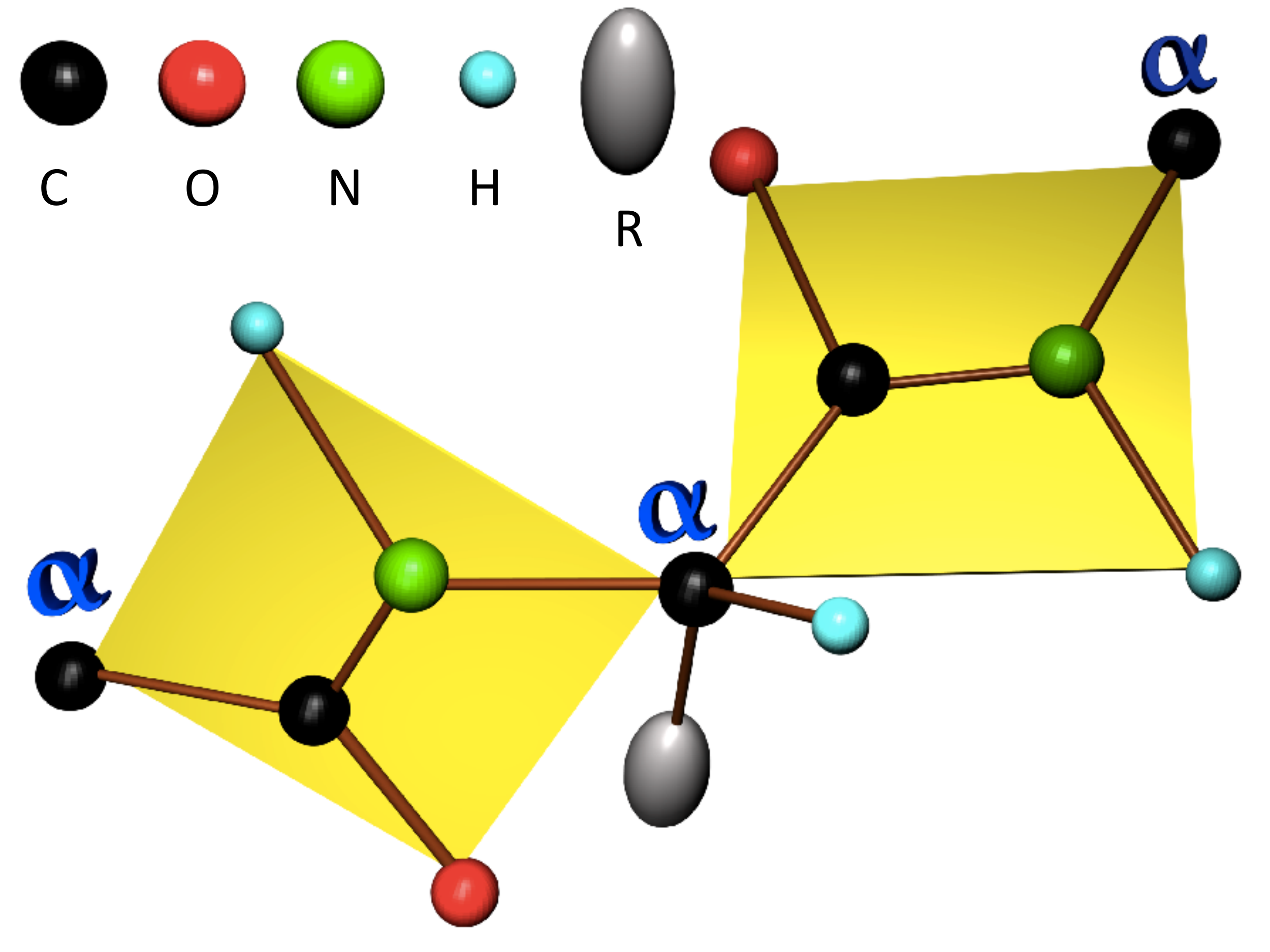}}
    \caption{All proteins are composed similarly, with an identical and {\it very} rigid peptide-plane 
    structure which makes Wilsonian universality operable: 
    The central C$\alpha$ carbons to which the twenty different amino acids (residues $R$) 
    are attached, form the vertices that connect the peptide planes into a one dimensional discrete string. 
    The distance between two consecutive C$\alpha$ atoms is 3.8 \AA, except in the rare case 
    of $cis$-proline where the distance 2.8 \AA~ should be used.
      } 
    \label{fig3}
  \end{center}
\end{figure}

However, proteins are not really continuous, differentiable strings.
Thus we need to extend our 
considerations accordingly. We start with the Frenet equations.
In the ''scaling limit'' where the concepts of Wilsonian universality become applicable,  
a protein is akin a  piecewise linear polygonal string. 
Its vertices coincide with 
the positions of the central C$\alpha$ carbons. See figure \ref{fig3}.
 {\it A priori} this reduction of the entire protein chain into a
skeletal C$\alpha$ backbone is an enormous simplification. 
But as we shall demonstrate, it is nevertheless sufficient for describing the structure 
and dynamics of a protein with a sub-atomic precision. The approach
we propose matches the accuracy 
which is obtained in ultrahigh resolution x-ray crystallography experiments.

Accordingly, we proceed to generalize the Frenet frame formalism to the case of polygonal strings 
that are piecewise linear. Let $\mathbf r_i$ be the vertices 
with $i=1,...,N$.  
At each vertex we introduce the unit tangent vector 
\begin{equation}
\mathbf t_i = \frac{ {\bf r}_{i+1} - {\bf r}_i  }{ |  {\bf r}_{i+1} - {\bf r}_i | }
\label{t}
\end{equation}
the unit binormal vector
\begin{equation}
\mathbf b_i = \frac{ {\mathbf t}_{i-1} - {\mathbf t}_i  }{  |  {\mathbf t}_{i-1} - {\mathbf t}_i  | }
\label{b}
\end{equation}
and the unit normal vector 
\begin{equation}
\mathbf n_i = \mathbf b_i \times \mathbf t_i
\label{n}
\end{equation}
The orthonormal triplet ($\mathbf n_i, \mathbf b_i , \mathbf t_i$) defines a
discrete version of the Frenet  frames  (\ref{curve})-(\ref{ncont}) 
at each position $\mathbf r_i$ along the 
string.

\begin{figure}[!hbtp]
  \begin{center}
    \resizebox{12.5cm}{!}{\includegraphics[]{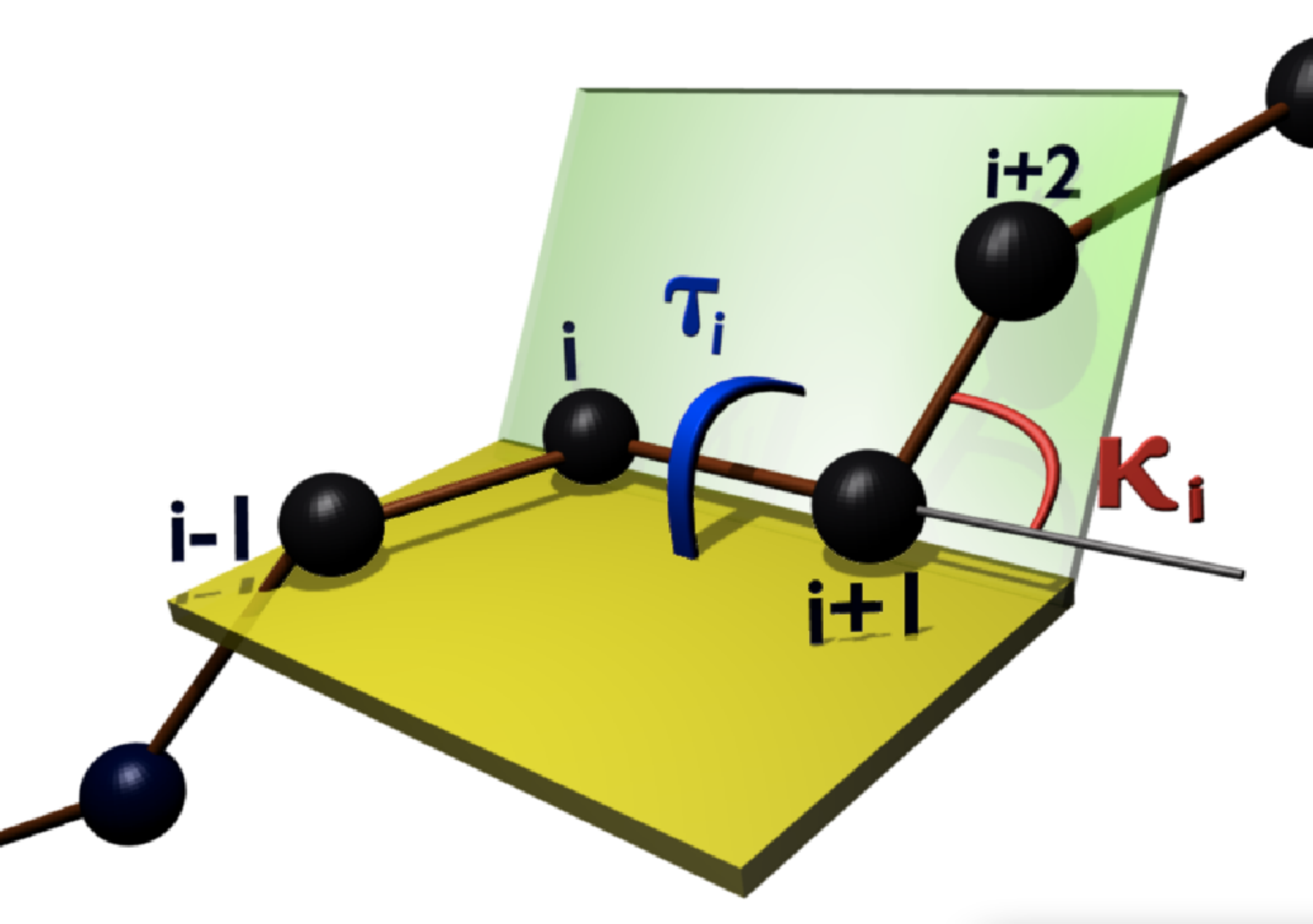}}
    \caption{Definition of bond ($\kappa_i$) and torsion ($\tau_i$) angles, along the discrete C$\alpha$ string.  } 
    \label{fig4}
  \end{center}
\end{figure}

In lieu of the curvature and torsion, we have the  bond angles and torsion angles, see figure \ref{fig4}.
The bond angles are
\begin{equation}
\kappa_{i} \ \equiv \ \kappa_{i+1 , i} \ = \ \arccos \left( {\bf t}_{i+1} \cdot {\bf t}_i \right)
\label{bond}
\end{equation}
and the torsion angles are
\begin{equation}
\tau_{i} \ \equiv \ \tau_{i+1,i} \ = \ {\rm sign}\{ \mathbf b_{i-1} \times \mathbf b_i \cdot \mathbf t_i \}
\cdot \arccos\left(  {\bf b}_{i+1} \cdot {\bf b}_i \right) 
\label{tors}
\end{equation}
When these angles are all known, we have the discrete Frenet equation
\begin{equation}
\left( \begin{matrix} {\bf n}_{i+1} \\  {\bf b }_{i+1} \\ {\bf t}_{i+1} \end{matrix} \right)
= 
\left( \begin{matrix} \cos\kappa \cos \tau & \cos\kappa \sin\tau & -\sin\kappa \\
-\sin\tau & \cos\tau & 0 \\
\sin\kappa \cos\tau & \sin\kappa \sin\tau & \cos\kappa \end{matrix}\right)_{\hskip -0.1cm i+1 , i}
\left( \begin{matrix} {\bf n}_{i} \\  {\bf b }_{i} \\ {\bf t}_{i} \end{matrix} \right) 
\label{DFE2}
\end{equation}
From this we obtain  the frame at position $i+i$ 
from the frame at position $i$. Once  all the frames have been constructed,  the entire string is obtained from
\begin{equation}
\mathbf r_k = \sum_{i=0}^{k-1} |\mathbf r_{i+1} - \mathbf r_i | \cdot \mathbf t_i
\label{dffe}
\end{equation}
Without any loss of generality we may set $\mathbf r_0 = 0$, choose $\mathbf t_0$ to 
point into the direction of the positive $z$-axis, and orient $\mathbf t_1$ to lie in the $y$-$z$ plane.

The relation (\ref{dffe}) does not involve the vectors $\mathbf n_i$ and $\mathbf b_i$.
In parallel with a continuum string, our discrete string remains intact under
rotations of  the ($\mathbf n_i, \mathbf b_i$) zweibein around $\mathbf t_i$.
Such a local SO(2)  rotation acts on the frames as follows
\begin{equation}
 \left( \begin{matrix}
{\bf n} \\ {\bf b} \\ {\bf t} \end{matrix} \right)_{\!i} \!
\rightarrow  \!  e^{\Delta_i T^3} \left( \begin{matrix}
{\bf n} \\ {\bf b} \\ {\bf t} \end{matrix} \right)_{\! i} =   \left( \begin{matrix}
\cos \Delta_i & \sin \Delta_i & 0 \\
- \sin \Delta_i & \cos \Delta_i & 0 \\ 
0 & 0 & 1  \end{matrix} \right) \left( \begin{matrix}
{\bf n} \\ {\bf b} \\ {\bf t} \end{matrix} \right)_{\! i}
\label{discso2}
\end{equation}
Here $\Delta_i$ is the rotation angle at vertex $i$ and we have introduced the SO(3) 
generators (\ref{T}). 
This yields the following transformation
of the bond and torsion angles, {\it cf.} (\ref{sokna}), (\ref{sotna}) 
\begin{equation}
\kappa_{i}  \ T^2  \ \to \  e^{\Delta_{i} T^3} ( \kappa_{i} T^2 )\,  e^{-\Delta_{i} T^3}
\label{sok}
\end{equation}
\begin{equation}
\tau_{i}  \ \to \ \tau_{i} + \Delta_{i-1} - \Delta_{i}
\label{sot}
\end{equation}
Since the $\mathbf t_i$ remain intact under (\ref{discso2}),
this transformation of ($\kappa_i, \tau_i$) has no effect on the discrete string geometry. 

{\it A priori}, the fundamental range of the bond angle is  $\kappa_i \in [0,\pi]$ while for the 
torsion angle the range is $\tau_i \in [-\pi, \pi)$. Thus we 
identify ($\kappa_i, \tau_i$) as the canonical 
latitude and longitude angles of a two-sphere $\mathbb S^2$. 
In parallel with the continuum case we find it useful to extend the range
of $\kappa_i$ into negative values $ \kappa_i \in [-\pi,\pi]$ $mod(2\pi)$. 
As in (\ref{kappaeta}) we compensate for this two-fold covering of $\mathbb S^2$ 
by the discrete $\mathbb Z_2$ symmetry
\begin{equation}
\begin{matrix}
\ \ \ \ \ \ \ \ \ \kappa_{k} & \to  &  - \ \kappa_{k} \ \ \ \hskip 1.0cm  {\rm for \ \ all} \ \  k \geq i \\
\ \ \ \ \ \ \ \ \ \tau_{i }  & \to &  \hskip -2.5cm \tau_{i} - \pi 
\end{matrix}
\label{dsgau}
\end{equation}
We note that this is a special case of (\ref{sok}), (\ref{sot}), with
\[
\begin{matrix} 
\Delta_{k} = \pi \hskip 1.0cm {\rm for} \ \ k \geq i+1 \\
\Delta_{k} = 0 \hskip 1.0cm {\rm for} \ \ k <  i+1 
\end{matrix}
\]

\section{discretized energy}
\label{sect12}

According to (\ref{DFE2}) the bond and torsion angles are the natural variables for 
constructing energy functions for discrete piecewise linear strings.
In analogy with the continuum case, the energy must remain invariant 
under the local SO(2) frame rotation (\ref{sok}), (\ref{sot}). 
We consider a generic energy function $H(\kappa,\tau)$ which is SO(2) invariant. 
We assume $H(\kappa,\tau)$ has an extremum with bond and torsion angle values
$ \kappa_i  = \kappa_{i0}$ and $ \tau_i  = \tau_{i0}$. We introduce a (small) deformation 
\[
\begin{matrix}
\Delta \kappa_i  & = &  \kappa_i - \kappa_{i0} \\
\Delta \tau_i  & = &  \tau_i - \tau_{i0}
\end{matrix}
\]
and we expand the energy around the extremum, 
\[
H(\kappa_i, \tau_i) \ = \ H(\kappa_{i0}, \tau_{i0}) + \sum\limits_k \{ \, 
\frac{\partial H}{\partial \kappa_k}_{|0}\! \Delta \kappa_k 
+ \frac{\partial H}{\partial \tau_k}_{|0}\! \Delta\tau_k \, \} 
\]
\begin{equation}
+  \sum\limits_{k,l} \left\{ \, \frac{1}{2}
\frac{\partial^2 H}{\partial \kappa_k \partial \kappa_l }_{|0}\! \Delta \kappa_k  \Delta \kappa_l
+ \frac{\partial^2 H}{\partial \kappa_k \tau_l}_{|0}\! \Delta\kappa_k \Delta \tau_l  +
\frac{1}{2}
\frac{\partial^2 H}{\partial \tau_k \partial \tau_l }_{|0}\! \Delta \tau_k  \Delta \tau_l
\, \right\} + \mathcal O (\Delta^3)
\label{Fexp}
\end{equation}
The first term evaluates the energy at the extremum. Since ($\kappa_{i0}, \tau_{i0}$) is an
extremum each term in the first sum vanishes. 
To proceed we bring to mind  that the energy and thus its expansion (\ref{Fexp}) only depends
on $\kappa_i$ and $\tau_i$ in a SO(2) gauge invariant manner.  
Accordingly, to the leading nontrivial order the 
expansion should coincide with the discretized version of the energy function (\ref{enenls}):
We rename ($\Delta \kappa,  \Delta \tau$) $\to$ ($\kappa, \tau$) that we identify with the geometrically
determined bond and torsion
angles defined in figure \ref{fig4}. Following the steps from (\ref{H1}) to (\ref{H2})  and up to an overall normalization factor we get
\begin{equation}
H(\kappa,\tau) =  \sum\limits_{i=1}^{N-1}  ( \kappa_{i+1}- \kappa_i)^2  + \sum\limits_{i=1}^N  
\left\{ \lambda\, (\kappa_i^2 - m^2)^2  
+ \frac{d}{2} \,  \kappa_i^2  \tau_i^2  - b \kappa_i^2 \tau_i
- a  \tau_i + \frac{c}{2} \tau_i^2\right\}
\label{A_energy}
\end{equation} 
For a detailed discussion of  (\ref{A_energy}) we refer to \cite{hu-2013,ioannidou-2014}.
Since the arguments that lead to (\ref{A_energy}) 
are based entirely on general symmetry principles that are universally valid, the result (\ref{A_energy}) 
is  {\it unique} for small fluctuations around a given background:
The  energy  (\ref{A_energy})
engages the complete set of gauge invariant quantities, in terms of
the bond and torsion angles, that emerge at leading order in 
a systematic Taylor expansion  of the full  energy around its local extremum.

The derivation of (\ref{A_energy}) utilizes only universal  principles. It describes
the structure and dynamics of {\it any} piecewise linear polygonal string, 
in the leading order of the fluctuations around the fixed background,   as
an extremum of the free energy.  In particular, in the case of proteins,
any energy function that describes the dynamics, either at all-atom level or at 
coarse-grained level,  {\it must}  reproduce (\ref{A_energy})
in the appropriate small fluctuation limit. 

For a complete treatment of Hamiltonian dynamics, the Poisson brackets of the variables ($\kappa_i,\tau_i$)
need to be determined. The brackets  that appear in the integrable DNLS hierarchy
\cite{faddeev-1987,ablowitz-2003} can be utilized.

\section{Discretized solitons}
\label{sect13}

The energy (\ref{A_energy}) is a variant of the energy that yields the discrete 
nonlinear Schr\"odinger (DNLS) equation \cite{faddeev-1987,ablowitz-2003,kevrekidis-2009}: 
The first term together with the $\lambda$ and $d$ 
dependent terms constitute the (naively) discretized Hamiltonian of the NLS model
in the Hasimoto variable.  The conventional DNLS equation is known to
support solitons. Thus we can try and find soliton solutions of (\ref{A_energy}). 

As in (\ref{taueq}) we 
first eliminate the torsion angle, 
\begin{equation}
\tau_i [\kappa] \ = \ \frac{ a + b\kappa_i^2}{c+d\kappa_i^2} \ = \ a \, \frac{ 1 + \hat b \kappa_i^2}{
\hat c + \hat d \kappa_i^2} \ \equiv a \hat \tau_i[\kappa]
\label{taueq2}
\end{equation}
For bond angles we then have
\begin{equation}
\kappa_{i+1} = 2\kappa_i - \kappa_{i-1} + \frac{ d V[\kappa]}{d\kappa_i^2} \kappa_i  \ \ \ \ \ (i=1,...,N)
\label{dnlse}
\end{equation}
where we set $\kappa_0 = \kappa_{N+1}=0$, and $V[\kappa]$ is given by (\ref{V}). This equation
is a deformation of the conventional  DNLS equation, it is not integrable  {\it a priori}.
For a numerical solution,  we convert  (\ref{dnlse})  into the following 
iterative equation \cite{molkenthin-2011,herr}
\begin{equation}
\kappa_i^{(n+1)} \! =  \kappa_i^{(n)} \! - \epsilon \left\{  \kappa_i^{(n)} V'[\kappa_i^{(n)}]  
- (\kappa^{(n)}_{i+1} - 2\kappa^{(n)}_i + \kappa^{(n)}_{i-1})\right\}
\label{ite}
\end{equation}
Here  $\{\kappa_i^{(n)}\}_{i\in N}$ denotes the $n^{th}$ iteration of an initial configuration  
$\{\kappa_i^{(0)}\}_{i\in N}$ and $\epsilon$ is some sufficiently small but otherwise arbitrary 
numerical constant; we often choose $\epsilon = 0.01$. 
The fixed point of (\ref{ite})  is clearly a solution of (\ref{dnlse}).
Once the numerically constructed fixed point is available, we calculate the corresponding 
torsion angles from (\ref{taueq2}). Then, we obtain the frames from (\ref{DFE2}) and 
proceed to construct the 
discrete string by (\ref{dffe}).

At the moment we do not know of an analytical expression of the soliton solution 
to the equation (\ref{dnlse}). 
But we have found \cite{chernodub-2010,hu-2011,krokhotin-2012}
that an {\it excellent} approximative solution can be obtained  by discretizing the topological 
soliton (\ref{kink}). 
\begin{equation}
\kappa_i \ \approx \   \frac{ 
m_{1}  \cdot e^{ c_{1} ( i-s) } - m_{2} \cdot e^{ - c_{2} ( i-s)}  }
{
e^{ c_{1} ( i-s) } + e^{ - c_{2} ( i-s)} }
\label{An1}
\end{equation}
Here ($c_1, c_2, ,m_{1},m_{2},s$) are parameters. The $m_{1}$ and $m_{2}$ 
specify the asymptotic $\kappa_i$-values of the soliton. Thus, 
these parameters are entirely determined by the character 
of the regular, constant bond and torsion angle structures that are adjacent to the 
soliton;  these parameters are not specific  to the soliton {\it per se}, but to the adjoining  
regular structures.
The parameter $s$ defines the location of the soliton along the 
string.  This leaves us with only two loop specific parameter, the $c_{1}$ and $c_{2}$. 
These parameters quantify the length of the bond angle profile that describes the soliton. 

For the torsion angle, (\ref{taueq2}) involves one parameter ($a$) that we have factored out as the overall relative scale
between the bond angle and torsion angle contributions to the energy. Then, there are three additional
parameters ($b/a, c/a, d/a$)  in the remainder $\hat \tau[\kappa]$. Two of these are again determined by the character 
of the regular structures that are adjacent to the soliton. 
As such, these parameters are not specific  to the soliton. The remaining single parameter
specifies the size of the regime where the torsion angle fluctuates.

The profile (\ref{An1}) is translation invariant. But on a lattice translation invariance 
is commonly broken by the Peierls-Nabarro barrier \cite{sieradzhan-2014}: When the soliton moves along the backbone 
lattice, quasi-particle waves are emitted in its wake. 
These waves drain the kinetic energy of the soliton, and cause it to decelerate. Eventually
the soliton becomes pinned to a particular backbone site and is unable to translate.

Once the soliton profile ($\kappa_i,\tau_i$) is known, we construct the ensuing discrete space string
from (\ref{DFE2}), (\ref{dffe}). In the case of a protein backbone  as shown in figures \ref{fig3}, \ref{fig4} 
where 
the vertices $\mathbf r_i$ coincide with the positions of the skeletal C$\alpha$
atoms,  we use a fixed value in (\ref{dffe}),
\[
|\mathbf r_{i+1} - \mathbf r_i |  \ = \ 3.8 \ {\rm \AA}
\]
 for the distance between neighboring vertices.
 The only exception is $cis$-proline in which case the distance 2.8 \AA~should be used; these are {\it very} rare. 
In our computations we shall also impose the steric constraint that prevents backbone self-crossing, in terms  
of the self-avoidance condition \cite{danielsson-2010}
\begin{equation}
|\mathbf r_{i} - \mathbf r_k | \ > \ 3.8 \ {\rm \AA} \ \ \ \ {\rm for \ } |i-k| \geq 2
\label{saw}
\end{equation}

On the regions adjacent to a soliton, 
we have constant values of $(\kappa_i, \tau_i)$.  In the case of a protein, these are the regions 
that correspond to the standard regular secondary structures.  For example, the 
standard $\alpha$-helix is
\begin{equation}
\alpha-{\rm helix:} \ \ \ \ \left\{ \begin{matrix} \kappa \approx \frac{\pi}{2}  \\ \tau \approx 1\end{matrix} \right.
\label{bc1}
\end{equation}
and the standard $\beta$-strand is 
\begin{equation}
\beta-{\rm strand:} \ \ \ \ \left\{ \begin{matrix} \kappa \approx 1 \\ \tau \approx \pi \end{matrix}  \right.
\label{bc2}
\end{equation}
Similarly, all the other familiar regular secondary structures such as 3/10 helices, 
left-handed helices {\it etc.} are described by definite constant values of $\kappa_i$ and $\tau_i$.
Protein loops are regions which are described by the soliton proper. The solitons
interpolate between the regular  structures, along a protein loop the values of ($\kappa_i, \tau_i$) 
are variable. 

Finally, in the case of a super-secondary structure 
(\ref{A_energy}) should be properly interpreted as the internal Landau
free energy, above the background of a regular secondary structure with 
constant values of $\kappa_i$ and $\tau_i$.

\section{Proteins and the Dirac problem of life}
\label{sect14}

Proteins are delicate nano-scale machines. Like other high precision 
machines, the way proteins function can be very sensitive 
to their conformation.  
The {\it protein folding problem} was originally posed some 50 years ago, and  it
has since then assumed various incarnations  \cite{dill-2008,dill-2012,pettitt-2013}.  
The problem  endures as one of the most important unresolved problems in science,
it aims to explain what is life.  
The {\it Sampo} would be a theoretical and computational
framework,  that predicts the shape and describes the
dynamics of a protein in its biological environment.  
The scale, complexity and trophy of the endeavor are enormous: There are some
25 million protein sequences that have been identified through
DNA sequencing. But thus far only around 100.000 structures have 
been experimentally determined \cite{berman-2000}.  Apparently, the average cost of a
structure determination by x-ray crystallography
is around 100 kUSD  and crystallization of a protein can sometimes take even years, if at all possible. 
We can hardly expect that the structures of a much larger percentage of sequences 
can ever be determined using the presently available experimental techniques. 
Therefore, the development of an accurate and reliable theoretical and computational
approach is pivotal for our ability to understand proteins, to resolve the problem of life. 

Moreover, a wrong fold is recognized as a common cause for a protein to loose its function. 
Misfolded proteins can be dangerous, even fatal, to a biological organism. For example 
neurodegenerative diseases like AlzheimerÕs and  ParkinsonÕs, 
diabetes-II, and many forms of cancer, are all caused by wrong folds in 
certain proteins. At the same time, the ability to elicit controlled protein misfolding 
might enable us to combat viral diseases like HIV and coronaviruses (SARS) where no
effective treatment presently exist.  Controlled protein misfolding might
eventually even open the door for the development of new generation molecular 
level antibacterials, to offset the emergence of resistance through evolutionary processes 
that are rapidly rendering many existing antibiotica  ineffective.  

 \vskip 0.2cm
\noindent 
There are, clearly, very many very good reasons to address the {\it Third Dirac Problem.}

\section{Yearning for the speed of life}
\label{sect15}

Among the goals of all-atom molecular dynamics (MD) is to provide an {\it ab-initio} description 
of protein folding and dynamics.  
MD utilizes finely tuned force fields like Charmm \cite{charmm} and 
Amber \cite{amber}  that  aspire to model all known (semi)classical
interactions between all the atoms, both in the protein and in the surrounding 
solvent (water).  
The Newtonian equations of motion are introduced, for each and every atom, including solvent. 
The equations are numerically integrated with a time step
around a femtosecond, which is the characteristic time scale of peptide bond vibrations.  
Using purpose built supercomputers like  {\it Anton} \cite{anton,anton2} and  
distributed computing projects like {\it folding$@$home} 
\cite{pande-2003}, the speediest MD simulations can reach a few 
micro-seconds of {\it in vivo} folding time, in a day 
{\it in silico} \cite{anton2}. This enables the modeling of relatively short  
and very fast folding proteins such as villin headpiece (HP35) and 
the $\lambda$-repressor protein (1LMB in Protein
Data Bank PDB \cite{berman-2000}), up to time scales that it takes  for these  proteins
to fold. 

But most proteins are considerable longer and fold during much longer time scales. 
For example, myoglobin which is a  protein described in all biochemistry 
textbooks, has 154 residues and folds in about 2.5 seconds \cite{myoglobin}. Thus, running 
at top speed of a few micro-seconds per day it should take {\it Anton} over 1.000 years  
to fold a myoglobin \cite{anton2}. And {\it Anton} is by far the fastest special-purpose MD 
machine ever built. Accordingly we need some $10^{11} $ orders of 
magnitude more in computer speed before MD simulations of proteins 
like myoglobin will take place {\it in silico} at the {\it speed of life}. 
Provided there is no compelling need to substantially increase the number of atoms 
involved.   Apparently the development of  processors with an ever increasing clock
speed has stalled.  As a consequence modeling of long time scale 
protein dynamics using all-atom molecular dynamics at the speed of life
does not appear realistic,  in the foreseeable future. 

Moreover, it does also appear that the presently available computers are incapable of 
handling sufficiently many individual atoms. As a consequence  
several essential all-atom ingredients still remain to 
be tackled by implicit and effective methods {\it in lieu} of all-atom. 
An important example is acidity, in case of explicit water. The  proper level of acidity
is pivotal to numerous  biological processes. 
Sometimes even a slight shift in acidity leads to a major change  in 
protein structure and function. A good example is amylin \cite{amylin}. It is  
a short polypeptide hormone 
that has been implicated in the onset of type-II diabetes. Much remains to be done to 
understand how amylin functions. In order to comprehensively mimic its 
properties computationally,  one needs to perform simulations  that model 
amylin both inside the $\beta$-cell 
granules of pancreas where pH $\approx$ 5.5, and in the extra-cellular domain 
with pH $\approx 7.4$ and
where the disease causing amyloidosis takes place; plus through the cell membrane. 
The structure of amylin seems to be
very different, in the two different pH environs. For an all-atom simulation to 
tackle the difference, suppose
we introduce $10^9$ explicit water molecules which is 4-5 orders of magnitude 
more than what  is presently possible, even with {\it Anton}. However,  
this only gives leeway 
for a mere few tens of explicit hydronium ions,  to model  
the physiological pH $\approx 7.4$ at all-atom level. This is hardly enough:
When it comes to acidity,  today's all-atom remains far 
from all atoms.

We dare to propose that, all-in-all, some 15 orders of magnitude increase, or even more, 
in the speed of computer simulations is needed before a copious and accurate all-atom description of 
protein folding dynamics at the speed of life becomes a reality. This is an enormous number, apparently 
comparable to or even exceeding 
the combined national debt of both USA and the  EU countries, in roubles.  From this 
perspective the protein folding problem sounds like doomed to endure among the pre-eminent 
unresolved conundrums in science, for a long time to  come. However, this enormous gap between 
the speed of life and the available speed {\it in silico} is also an excellent opportunity: 
There is a vast {\it Every Man's Land} available for the development of alternative, computationally 
effective approaches. 
For this purpose, 
various coarse-grained models are being developed.
Contributions to the all-atom 
force fields  that are presumed  to be less relevant, 
are systematically eliminated. A simplified geometry can also  be used. 
For example, the UNRES \cite{liwo} force field provides a very detailed and finely tuned coarse 
grained potential energy with some 15 different terms, in combination with a simplified geometry. 
Present coarse-graining can extend all-atom MD simulations 
by some 4-5 orders of magnitude.
This considerable success of coarse-graining raises the question, what are the truly relevant 
contributions to the free energy function. 

We note that there are also highly simplified approaches such as the G\=o 
model \cite{gomodel}  and its variants.  In a G\=o-type model one
constructs the energy function from the knowledge of 
all atomic positions and native contacts in the protein 
of interest: There are as many, or even more parameters  than there are atomic positions.
As a consequence these approaches 
do not have any predictive power for the native fold. 
But they can still be profited {\it e.g.} to study how the folding might 
proceed.

In addition, there are several highly successful non-Physics based approaches 
to the protein folding problem:
Structural classification schemes \cite{cath,scop} 
reveal that despite enormous diversity in the amino acid edifice, 
the number of different folds observed in PDB structures
is quite small.  This empirical observation forms the conceptual basis 
for {\it de novo}  structure prediction methods \cite{template}. 
The basic idea is to utilize properly chosen fragments that are found in the protein conformations 
which have already been deposited  in the PDB database, 
as modular building blocks very much like {\it Lego bricks} to construct a folded protein. 
These comparative methods have the best predictive power \cite{casp} for crystallographic folded protein structure, 
at the moment. However, the lack of a well grounded energy function 
impedes their utility in investigations of dynamical aspects  like  the way how the
folding takes place.

Finally, what to some might appear as a sign of desperation, we mention  
and recommend the  on-line 
gaming experience  {\it Foldit}
\[
{\tt http://fold.it}
\]
The players help scientists to discover  and determine how a given protein might fold. Best players display
an  impressive ability to do so, they sometimes even  beat scientific approaches.

\section{Thermostats}
\label{sect15a}

Among the theoretical issues where we trust important progress 
remains to be made,  is the development of  thermostats \cite{thermo}. 
Molecular dynamics integrates the all-atom Newton's equations 
of motion, hence  the total energy is a conserved quantity. As a consequence 
MD describes a protein in a microcanonical ensemble. But this does not correspond 
to {\it in vivo}: The biological environ of a protein is characterized by a constant 
temperature,  the natural habitat of life in balance is in a canonical ensemble.
A native state of a protein which is in (local)  thermodynamical equilibrium with its environ 
corresponds to a local minimum of the Helmholtz free energy 
\begin{equation}
H \ = \ E - TS 
\label{helm}
\end{equation}
where $E$ is the appropriate internal energy, $T$ is temperature and $S$ is the entropy; we overlook volume effect.
Note that the entropic forces that derive from $S$ are pivotal for a protein collapse.  
In order to describe this {\it in vivo} environment, the all-atom Hamiltonian Newton's 
equations need to be modified to mimic a constant  temperature setting. For this purpose
diverse thermostats have been  developed. They facilitate the MD simulation 
of protein folding, mostly under stationary non-equilibrium conditions. 
It remains a delicate challenge to construct a purely Hamiltonian thermostat that both models the
canonical ensemble of the original system, and allows for a computationally effective
discretization for speedy reliable simulations.  

Thermostatting often involves 
a deformation of the all-atom Newton's equation into a Langevin equation. However, such
an equation is both non-deterministic and lacks time-irreversibility, and the original 
impetus to consider a Hamiltonian framework becomes lost. Unfortunately, it appears 
impossible to construct a canonical thermostat 
Hamiltonian that has both a finite number of thermostat variables and a non-singular potential.
Maybe the most widely used canonical thermostat in MD simulations of protein folding is 
the one by Nos\'e and Hoover \cite{nose,hoover}. 
In its simplest variant the thermostat has a Hamiltonian character, albeit with a singularity. 
The all-atom phase space is extended by a single {\it ghost} particle with a logarithmically
divergent potential, its r\^ole is to provide temperature for all the rest. 

We consider a thermostatted  extension of  (\ref{swave})
\begin{equation}
S \ = \ \int\limits_{-\infty}^\infty \! dt \, \left\{  \frac{1}{2} q^2 \kappa_t^2 + V[\kappa] + \frac{1}{2} q_t^2 + T
\ln q \right\}
\label{nose1}
\end{equation}
We  assume that $V[\kappa]$ has a double-well 
profile.  When  $q(t)\equiv 1$ we interpret $S$ as the Euclidean action of  
$\kappa$;  The variable $q$ is akin the Nos\'e-Hoover 
thermostat, with  $\kappa$ a single generic coordinate of the thermostatted system.

A finite value of (\ref{nose1}) is imperative so that the semi-classical 
amplitude of  the coordinate $\kappa$ to cross over the potential barrier between
the two distinct minima of $V[\kappa]$, is finite.
We are interested in the effect of the thermostat $q$ on the barrier crossing amplitude.
The equations of motion are
\begin{eqnarray}
q^2 \kappa_{tt}  & = &  V_\kappa - 2 q \, q_t \kappa_t \  \simeq  V_\kappa - \gamma \kappa_t
\label{nose2a}
\\
q\, q_{tt} & = & q^2 \kappa_t^2  + T 
\label{nose2b}
\end{eqnarray}
Note that the coupling between $\kappa$ and the thermostat variable  gives rise to an effective friction-like 
coefficient $\gamma(t)$. In the absence of $q$, we assume the equation (\ref{nose2a})  supports 
a finite action instanton, {\it e.g.} with a profile that resembles (\ref{kink}).
We inquire whether  (\ref{nose1}) continues to support a finite action 
instanton, in the presence of the thermostat. 

Suppose
\[
\kappa(t) \ \buildrel{t\to\pm\infty}\over{\longrightarrow} \ \kappa_\pm
\]
where $\kappa_\pm$ are values of the coordinate $\kappa$ at the opposite sides of the potential barrier.
Then, for finite action we obtain the Gibbsian format
\begin{equation}
q(t) \ \buildrel{t\to\pm\infty}\over{\longrightarrow} \ q_\pm \ = \ e^{-\frac{1}{T}  V[\kappa_\pm]}
\label{gibbsi}
\end{equation}
This  proposes that $T$ is like temperature, when positive valued.
We integrate (\ref{nose2b}), 
\[
\int\limits_{-\infty}^\infty \! dt \, \left\{  q_t^2 + q^2 \kappa_t^2 \right\} \ = \ -  
\int\limits_{-\infty}^\infty \! dt \,  T
\]
For a finite  Euclidean 
action (\ref{nose1}),  the integral on the {\it l.h.s.} is  finite. Since it is non-negative, 
the temperature $T$, if indeed positive,  must vanish. 
For finite  $S$, we deduce that $q(t)$ can not be viewed as a variable that yields an equation of motion,  it is merely a 
fixed background field with a fixed profile and no dynamics:
For an instanton to persist  any dynamical 
thermostat field $q(t)$ should become {\it entirely} decoupled. 

We conclude that thermostatting should be 
performed  with due care. Impetuous thermostatting 
can disfigure  the non-perturbative structure. Even  to 
the extent that soliton-like configurations entirely disappear.  This would be unfortunate:
We proceed to argue that topological solitons are the modular building 
blocks of folded proteins.

\section{Solitons and proteins}
\label{sect16}

Various taxonomy schemes such as CATH and SCOP  \cite{cath,scop}  
reveal that  folded proteins have a modular build. Novel
topologies are rare, to the extent that some authors think most
modular building blocks  are already known 
\cite{rackovsky-1990,skolnick-2009}. This convergence in protein architecture 
is a palpable manifestation that protein folding is driven by a universal structural 
self-organization principle. 

We argue  that a DNLS soliton is the {\it 
auriga praecipua}. Indeed, it has  been shown
that over 92$\%$ of all C$\alpha$-traces of PDB proteins can be
described by 200 different parametrizations of the discretized NLS kink (\ref{An1}), 
with better than 0.5 \AA~ root-mean-square-distance (RMSD) precision \cite{krokhotin-2012}.
Accordingly, we set up to describe  the modular building blocks of proteins in terms of
various parametrizations of the DNLS soliton profile, that is described by the 
equations (\ref{ite}), (\ref{taueq2}), (\ref{DFE2}) and (\ref{dffe}). 

\begin{figure}[!hbtp]
  \begin{center}
    \resizebox{12.5cm}{!}{\includegraphics[]{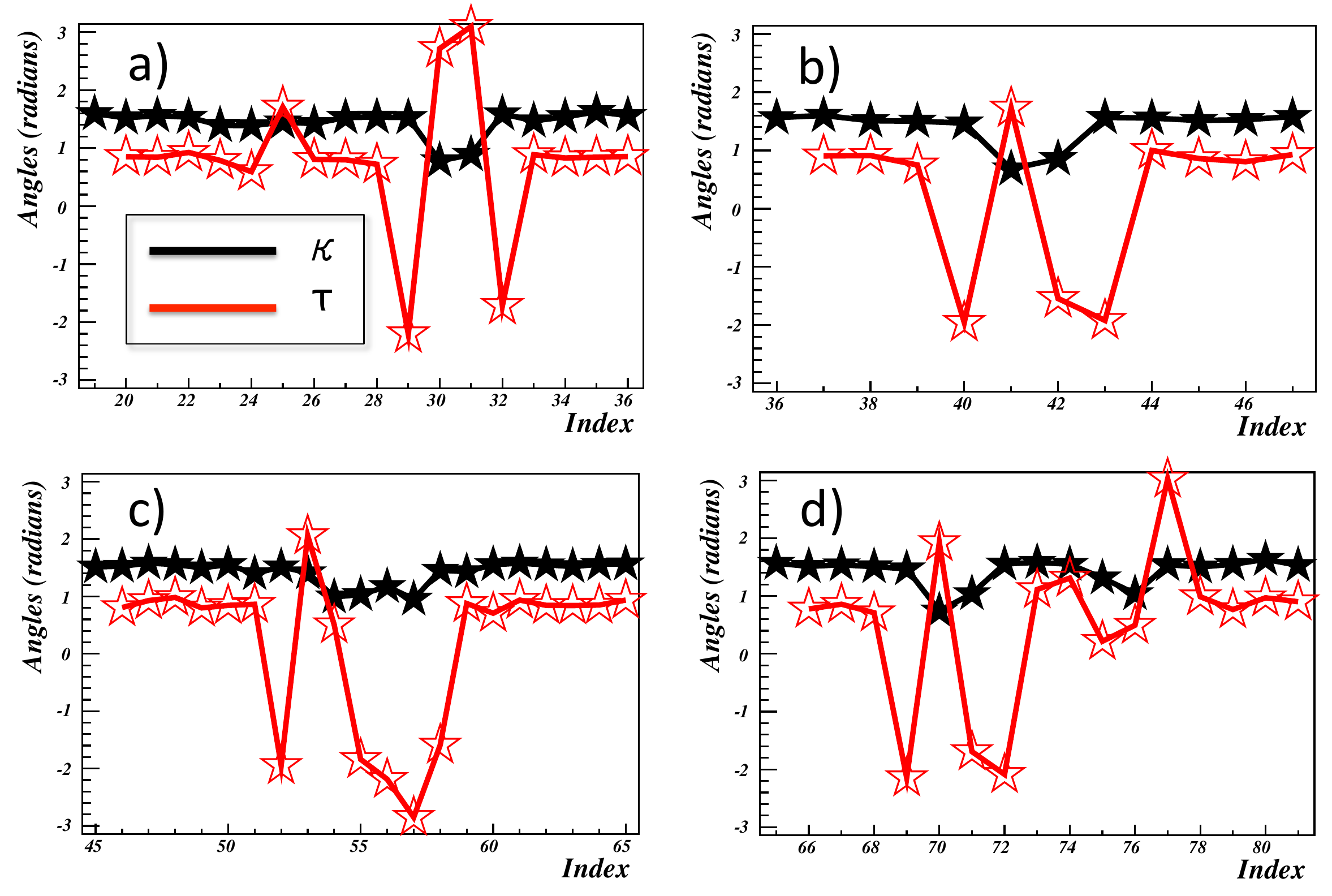}}
    \caption{The bond  angle ($\kappa$) and torsion angle ($\tau$) spectrum of $\lambda$-repressor 1LMB; the indexing follows PDB. } 
    \label{fig5}
  \end{center}
\end{figure}

From the C$\alpha$ coordinates of a given protein, available at PDB, 
we compute the backbone bond and torsion angles. For this we {\it initially}  fix the $\mathbb Z_2$ 
gauge in (\ref{dsgau}) so that all the bond angles take positive values.  
A generic profile consists of a set of $\kappa_i$, typically between 
$\kappa_i \approx 1$ and $\kappa_i \approx \pi/2$ and
the upper bound is due to steric constraints.  
The torsion angle values $\tau_i$ are much more unsettled, they
jump over the entire range  from $-\pi$ to $+\pi$. 
In figure \ref{fig5} we show as an example the ($\kappa_i,\tau_i$) spectrum in the case 
of the $\lambda$-repressor protein, with PDB code 1LMB. 
The spectrum is fairly typical, for a PDB configuration.

\begin{figure}[!hbtp]
  \begin{center}
    \resizebox{12.5cm}{!}{\includegraphics[]{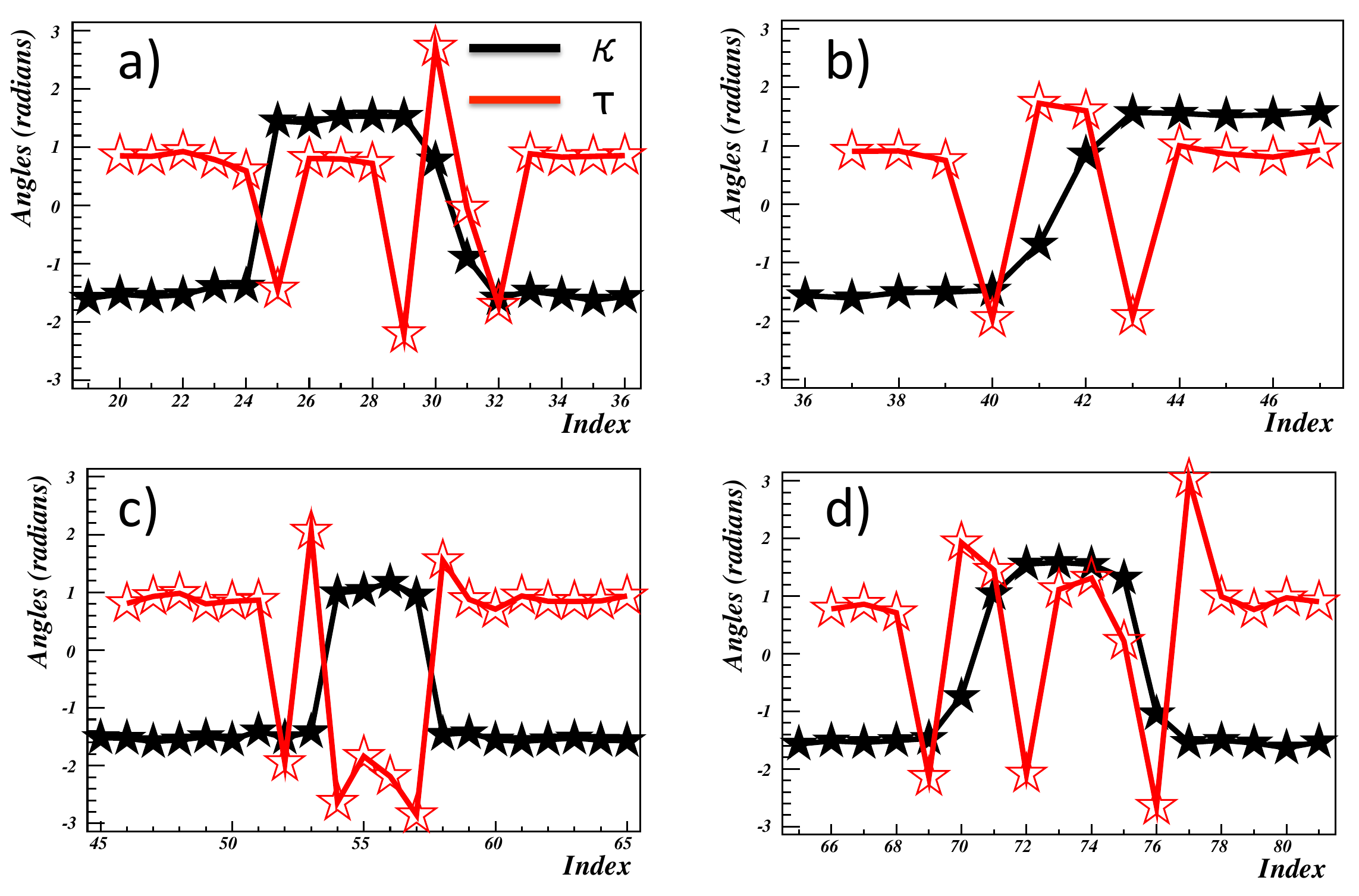}}
    \caption{The $\mathbb Z_2$ gauge transformations of the bond  angle ($\kappa$) and torsion angle ($\tau$) 
    spectrum of $\lambda$-repressor 1LMB; the indexing follows PDB. } 
    \label{fig6}
  \end{center}
\end{figure}

The C$\alpha$ backbone of a protein is  piecewise linear, the 
spectrum of ($\kappa_i, \tau_i$) is discrete.  The
general bifurcation analysis in Sections \ref{sect10} and \ref{sect10a} relates 
to a continuous string, with differentiable 
curvature and torsion. Thus we need an extension of the general results to the specific case 
of a piecewise linear string:  We continue to interpret
a change in the sign of $\tau_i$ in terms  of a flattening point. It suggests that an inflection
point perestroika has taken place. Accordingly, we implement a series of 
$\mathbb Z_2$ gauge transformations (\ref{dsgau}) in the vicinity of putative flattening points 
where $\tau_i$ changes sign or is otherwise
unsettled, to identify the putative multi-soliton profile in $\kappa_i$. 
For example, in the case of 1LMB there 
are four regions with an irregular $\tau_i$
profile. By a judicious choice of $\mathbb Z_2$ gauge transformations we identify 
seven different solitons (\ref{An1}) in $\kappa_i$.  
The profiles are shown in figure \ref{fig6}. 
Each of the
soliton profile is clearly accompanied by putative flattening points; note the multivaluedness  of
$\tau_i$. The general considerations in Sections \ref{sect10} and \ref{sect10a}, albeit developed for 
the case of continuous strings, are very much in line with the analysis of a generic discrete C$\alpha$ protein
profile. We conclude that protein folding is due to inflection and flattening point perestroika's. These bifurcations 
deform the C$\alpha$ backbone and create DNLS solitons along it.

\begin{figure}[!hbtp]
  \begin{center}
    \resizebox{12.5cm}{!}{\includegraphics[]{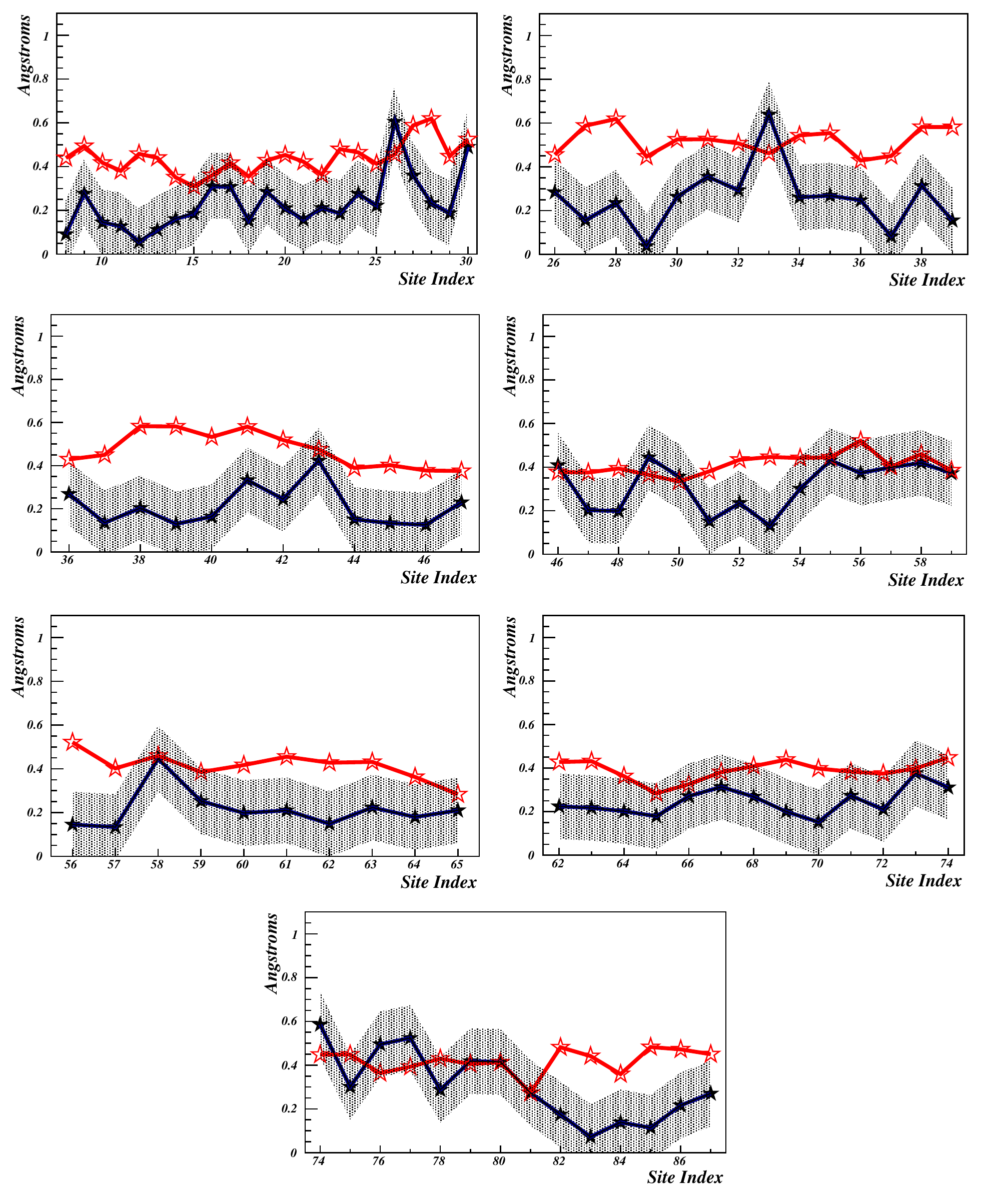}}
    \caption{The distance between the PDB backbone of the first 1LMB chain and its approximation by the seven solitons solutions. 
    The black line denotes the distance between the soliton and the corresponding PDB 
    configuration. The grey area around the black line describes our estimate of 15 pico meter (essentially 
    quantum mechanical) zero point fluctuation 
    distance around each soliton. The grey (red) line denotes the Debye-Waller fluctuation distance (\ref{Bfac}). } 
    \label{fig7}
  \end{center}
\end{figure}

In the case of our example 1LMB, the seven 
$\mathbb Z_2$ gauge transformed soliton profiles define the background, around which
we perform the expansion (\ref{Fexp}), (\ref{A_energy}). For this we first train 
the energy function to describe the background. In practice we do the training by
demanding that the fixed point of the iterative 
equation (\ref{ite}) models the C$\alpha$ 
backbone as a DNLS multi-soliton solution, and
with a prescribed precision.  We have developed 
a program {\it GaugeIT} that implements the $\mathbb Z_2$ gauge transformations
to identify the background, and we have developed a program {\it PropoUI} to train the energy so that
its extremum  models the background as a multi-soliton. 
The programs are described at 
\begin{equation}
{\tt http://www.folding-protein.org}
\label{propro}
\end{equation}
In the case of a protein for which the PDB structure is determined with an ultra-high resolution,
typically below 1.0 \AA ngstr\"om, {\it PropoUI} routinely constructs
a multi-soliton that describes the C$\alpha$ backbone with the
experimental precision: The accuracy of a given experimental PDB structure can be 
estimated from  the B-factors using  
the Debye-Waller formula. It relates the experimental  B-factor to the  one standard deviation 
fluctuation distance in the C$\alpha$ position 
\begin{equation}
\sqrt{<\mathbf x^2>} \ \approx  \ \sqrt{\frac{B}{8\pi^2}}
\label{Bfac}
\end{equation}
The B-factors are available in PDB. In figure  \ref{fig7} 
we compare the distance between the
C$\alpha$ backbone, the DNLS soliton solution, and the B-factor fluctuation distance in the 
experimental structure 1LMB. 
As shown in the figure, the DNLS soliton describes the backbone with a precision that is fully
comparable with the experimental uncertainties. The grey zone around the soliton profile denotes our
best estimate for the extent of quantum mechanical zero-point fluctuations. By
analyzing available crystallographic PDB structures we have concluded that
the quantum mechanical fluctuations in the positions of the C$\alpha$ atoms should not exceed 
15 pico meters. This  estimate coincides with the historically used value, for the wavelength 
boundary between x-rays and $\gamma$-rays. 
 
 \vskip 0.2cm

Simulations that have been performed using the UNRES force field, thus far,  support that folded proteins display 
a soliton driven structural self-organization. Furthermore, the cause of  protein folding can be traced to 
a combination of inflection point and bi-flattening perestroika's.  At least in the case of protein-A \cite{maisu}.

\section{Folding at the speed of life}
\label{sect17}

By construction, the  expression (\ref{A_energy}) of the energy  is {\it universal}:
It is the leading infrared contribution to the expansion (\ref{Fexp}) of the full Helmholtz free energy (\ref{helm}), 
around a generic but pre-determined extremum. Thus  (\ref{A_energy})
describes  the energy landscape of a protein, not only at the extremum but also  in its
vicinity.  The expansion (\ref{A_energy}) 
can be utilized to explore the near-equilibrium dynamics, 
such as the way how the protein responses to temperature fluctuations 
and variations in other environmental 
parameters, acidity and so forth.  

In particular,  the present approach is designed to facilitate the description of  protein 
dynamics over biologically
relevant time scales. It  averages over all very short time scale 
atomic level oscillations, vibrations, and those tiny fluctuations and deformations in the positions 
of the individual atoms that are more or less irrelevant to the way how the 
folding progresses over time scales that are  biologically important. 

To describe non-equilibrium dynamics, we adopt a Markovian Monte Carlo (MC) time evolution with 
the universal heat bath probability
distribution  (Glauber dynamics) \cite{glauber,lebo}
\begin{equation}
\mathcal P = \frac{x}{1+x} \ \  \ \ {\rm with}  \   \ \ \ x =     \exp\{ - \frac{ \Delta E}{kT} \}  
\label{P}
\end{equation}
Here $\Delta E$ is the energy difference between consecutive MC time steps,  that we 
compute from (\ref{A_energy}). It can be proven \cite{,marti1,marti2} that MC simulation 
with (\ref{P}) approaches the Gibbsian distribution at exponential rate. Thus, the method
should provide a statistically meaningful description of near-equilibrium protein folding
dynamics, at least during adiabatic temperature variations.

In our simulations of near-equilibrium proteins, we renormalize  the 
numerical value of the temperature factor $kT$ so that it coincides 
with the experimentally observed $\theta$-point temperature \cite{omagla}.
We proceed as  follows:  We start by training (\ref{Fexp}), (\ref{A_energy}) 
to describe a given, typically very low temperature crystallographic  PDB protein configuration as 
a multi-soliton.
For this we utilize the program {\it Propro} that we have described in (\ref{propro}).
Once the multi-soliton has been constructed, we subject it to extensive heating and cooling 
simulations. We  start
from a vanishingly low temperature value, with no apparent thermal fluctuations 
in the C$\alpha$ positions.  We slowly increase the temperature until we observe a structural 
transition akin a phase transition, above which the configuration resembles a random walker.
The transition identifies  the renormalization point of the temperature factor,  it takes place at
 the $\theta$-point temperature. A convenient order parameter for detecting the 
 $\theta$-point  is the C$\alpha$-trace 
radius of gyration \cite{degennes,schafer}
\begin{equation}
R^2_g \ = \ \frac{1}{2N^2}  \sum_{i,j} ( {\bf r}_i  
- {\bf r}_j )^2 \  \buildrel{N \ {\rm large}}\over{\longrightarrow}  \
R_0^2 N^{2\nu} 
\label{R0}
\end{equation}
Here $\nu$  is the compactness index that governs  the large-$N$ asymptotic form of equation (\ref{R0}), 
and $R_0$ is a form factor that characterizes the effective distance between the C$\alpha$ atoms
in the large $N$  limit. The compactness index  $\nu$ is a universal quantity but the form factor $R_0$ is not.
The form factor is in principle a calculable quantity, 
from the atomic level structure of the protein and the surrounding solvent.

\begin{figure}[!hbtp]
  \begin{center}
    \resizebox{12.5cm}{!}{\includegraphics[]{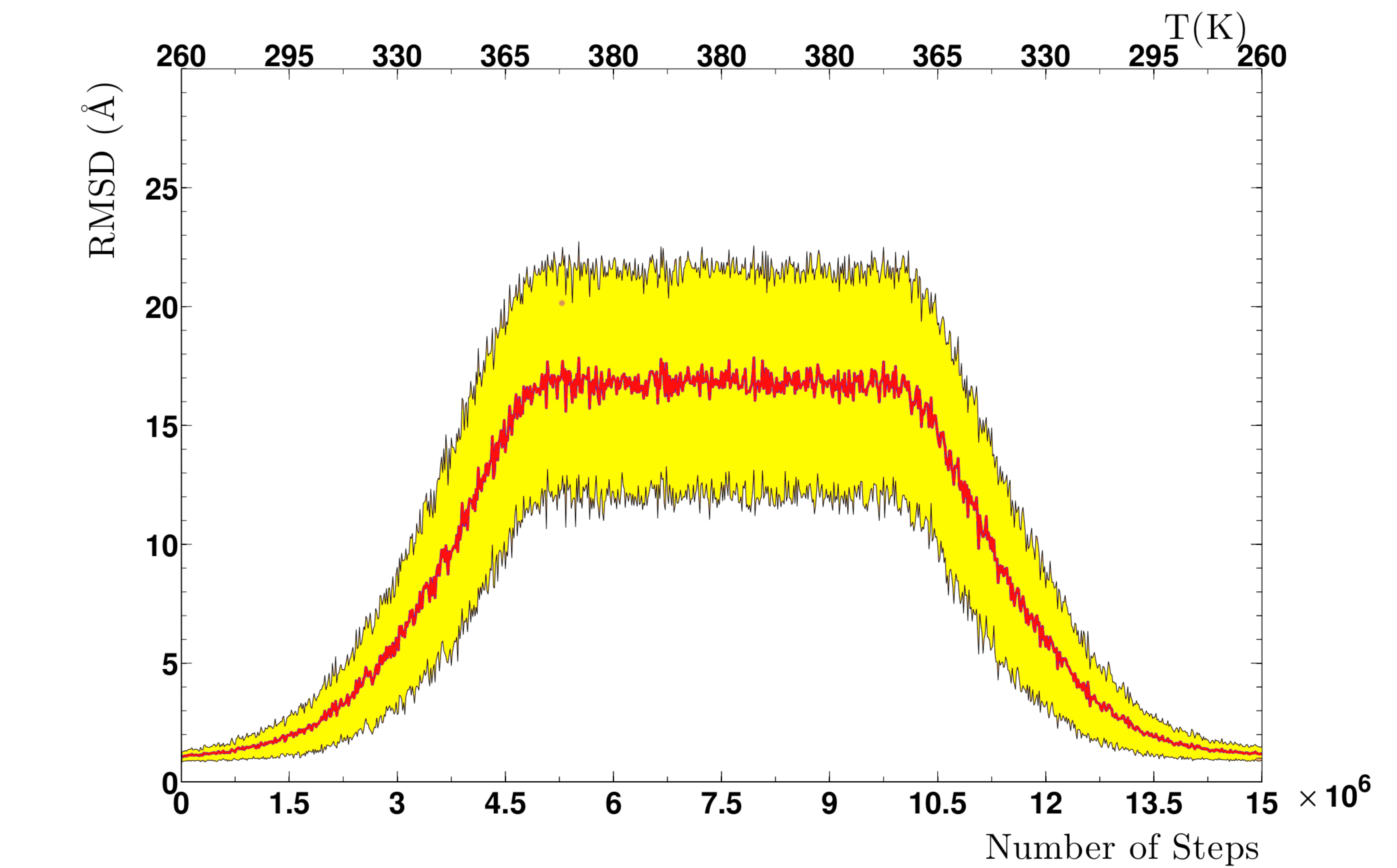}}
    \caption{The evolution of the root-mean-square distance (RMSD) 
    between the myoglobin PDB entry 1ABS backbone,  and the simulated 
    multi-soliton configuration. The (red) line is the average of 1.000 simulations, and the surrounding (yellow) shaded area 
    describes the one standard deviation extent of fluctuations. Along the top axis, we have converted the temperature 
    into Kelvin scale, using the renormalization
    procedure described in \cite{myoglo}.
   } 
    \label{fig8}
  \end{center}
\end{figure}

In figure \ref{fig8} we show, as an example, how the C$\alpha$ root-mean-square-distance (RMSD) between 
the crystallographic X-ray myoglobin structure with PDB entry code 1ABS and its multi-soliton 
description constructed using  (\ref{Fexp}), (\ref{A_energy}) 
evolves, when we increase and decrease the temperature.
\noindent
In this simulation we first heat up the multi-soliton. We thermalize it in the $\theta$-regime. We then 
cool it down,  back to low temperature values 
where only very small thermal motion persist.

Generically, in the case of proteins with a well defined 
native state such as myoglobin and $\lambda$-repressor, both of which  we have already introduced  as examples,
we recover the initial configuration with a very high RMSD precision as shown in figure \ref{fig8}. 
But for a protein that is intrinsically disordered, an example is
amylin that we described in Section \ref{sect15}, this is not the case. For an intrinsically disordered
protein we commonly find  a complex conformational landscape in the low temperature limit.

In the case 
of myoglobin, when we start from a random configuration 
above the $\theta$-point temperature, say $380K$ in figure \ref{fig8},
we reach the native state  in over 99 per cent of
cooling simulations within 3.5 seconds using a single processor in MacBook Air.  
This can be contrasted to the experimentally measured {\it in vivo} folding time 
which is 2.5 seconds \cite{myoglobin}. Thus,  by describing proteins as  
multi-solitons it is  quite possible to reach the  {\it speed of life} with a presently 
available, standard laptop. Even to exceed it, with a good work-station.

\section{Concluding remark}

The collapse of a protein is a complex physical phenomenon that engages a multitude of disparate temporal 
and spatial scales. In particular, there are many high energy barriers that the protein must be able to
overcome as it progresses from a random string towards the native fold. This obligates
thousands of atoms to cooperate over quite long time periods, so that complex
collective multi-molecule  motions can take place and enable the conformation to cross 
the various steep hurdles and hindrances. These obstacles that come with varying  scales 
and diverse structures, pose major computational bottle-necks in any all-atom approach 
to the protein folding problem. Thus, a detailed MD simulation of the entire 
folding process remains a formidable  task. But in a seemingly 
paradoxical manner \cite{levinthal} proteins succeed to fold 
in the congestion of our cells, very reliably and at a  quite high speed.  

Similar kind of apparent paradoxes are encountered all over the physical world: 
Think of water, how quickly it finds a way to self-organize into a wave. 
Or a typhoon, that commonly emerges in the atmosphere. Each involve the collective cooperation of 
an enormous number of atomic level constituents, far more 
than in a protein. Any attempt to an all-atom description would be preposterous. 
But in each case like in numerous others, we have an exceedingly solid 
theoretical framework: Korteweg-de Vries equation
with its solitons in the former and the vortices of Navier-Stokes equation in the latter.  
Why not try and use a similar kind of approach to structural self-organization, also 
in the case of proteins?

\section{Acknowledgements:}
I am indebted to Ludvig Faddeev for our continuing friendship and scientific collaboration that spans over 25 years. 
Most of the results presented  in this article are based at least indirectly on our joint work, all I do is describe 
what I have learned by working with him. Ludvig also knows the art of
a rich life, and how to  enjoy it at full speed. Sometimes, only 
a feather on the brake (see acknowledgement of reference \citenum{faddeev-1999-b}).
I wish to  thank S.Chen, M.Chernodub, U. Danielsson, M.-L. Ge, J. He, K. Hinsen, 
Y. Hou, S. Hu,  N. Ilieva, T. Ioannidou, Y. Jiang,  D. Jin, G. Kneller,
A.Krokhotin, J.Liu, A. Liwo, M.Lundgren,   G. Maisuradze, 
N.Molkenthin, X. Nguyen, S. Nicolis, X. Peng, H. Scheraga, F. Sha, A. Sieradzan,  A. Sinelkova, M. Ulybyshev 
and many many others for numerous discussions on various aspects of the research  described here.
I acknowledge support from Region Centre 
Rech\-erche d$^{\prime}$Initiative Academique grant, Sino-French 
Cai Yuanpei Exchange Program (Partenariat Hubert Curien), 
Vetenskapsr\aa det, Carl Trygger's Stiftelse f\"or vetenskaplig forskning, and  Qian Ren Grant at BIT.

\end{document}